\documentclass{elsart}
\usepackage{graphicx}
\usepackage{amssymb}
\begin{document}

\begin{frontmatter}
\title{Analysis of the low-energy $\pi^\pm p$ elastic-scattering data}
\author[EM]{E. Matsinos{$^*$}},
\author[GR]{G. Rasche},
\address[EM]{Centre for Applied Mathematics and Physics, Zurich University of Applied Sciences, Technikumstrasse 9, P.O. Box, CH-8401 Winterthur, Switzerland}
\address[GR]{Institut f\"{u}r Theoretische Physik der Universit\"{a}t, Winterthurerstrasse 190, CH-8057 Z\"{u}rich, Switzerland}

\begin{abstract}
We report the results of a phase-shift analysis (PSA) of the low-energy $\pi^\pm p$ elastic-scattering data. Following the method which we had set forth in our previous PSA \cite{mworg}, we 
first investigate the self-consistency of the low-energy $\pi^\pm p$ elastic-scattering databases, via two separate analyses of (first) the $\pi^+ p$ and (subsequently) the $\pi^- p$ 
elastic-scattering data. There are two main differences to our previous PSA: a) we now perform only \emph{one} test for the acceptance of each data set (based on its contribution to the overall 
$\chi^2$) and b) we adopt a more stringent acceptance criterion in the statistical tests. We show 
that it is possible to obtain self-consistent databases after removing a very small amount of the data ($4.57 \%$ of the initial database). We subsequently fit the ETH model \cite{glmbg} to 
the truncated $\pi^\pm p$ elastic-scattering databases. The model-parameter values show reasonable stability when subjected to different criteria for the rejection of single data points and 
entire data sets. Our result for the pseudovector $\pi N N$ coupling constant is $0.0726 \pm 0.0014$. We extract the scattering lengths and volumes, as well as the $s$- and $p$-wave hadronic 
phase shifts up to $T=100$ MeV. Large differences in the $s$-wave part of the interaction can be seen when comparing our hadronic phase shifts with the current SAID solution (WI08); there is 
general agreement in the $p$ waves, save for the $\tilde{\delta}_{1-}^{1/2}$ hadronic phase shift.\\
\noindent {\it PACS:} 13.75.Gx; 25.80.Dj
\end{abstract}
\begin{keyword} $\pi N$ elastic scattering; $\pi N$ hadronic phase shifts; $\pi N$ coupling constants; $\pi N$ threshold constants
\end{keyword}
{$^*$}{Corresponding author. E-mail: evangelos.matsinos@zhaw.ch, evangelos.matsinos@sunrise.ch; Tel.: +41 58 9347882; Fax: +41 58 9357306}
\end{frontmatter}

\section{\label{sec:Introduction}Introduction}

This is the first of three papers dealing with issues of the pion-nucleon ($\pi N$) interaction at low energies (pion laboratory kinetic energy $T \leq 100$ MeV). The goal in this study 
is to update part of the material given in our previous phase-shift analysis (PSA) \cite{mworg}. Hereafter, we will refer to that solution as UZH06, and to the one obtained in the present 
work as ZUAS12. We will show that the differences between these two solutions are small. In the second of the papers \cite{mrw2}, we will address the self-consistency of the only $\pi^\pm p$ 
elastic-scattering data which appeared in the meantime, i.e., of the measurements of Denz \etal~(DENZ04) \cite{denz}, give details on the problems we encountered in the analysis of these 
measurements, and provide evidence to support our decision to retain the UZH06 initial $\pi^\pm p$ elastic-scattering databases. In the third paper \cite{mrw3}, we will analyse all 
available experimental data for the charge-exchange (CX) reaction $\pi^- p \rightarrow \pi^0 n$ and re-address the violation of the isospin invariance in the hadronic part of the 
$\pi N$ interaction. In our program, we make use of the electromagnetic (em) corrections of Refs.~\cite{gmorw}, which lead to the em-modified hadronic quantities \cite{mworg}.

We re-open the case of the $\pi N$ interaction at low energies because of three reasons.
\begin{itemize}
\item There have been changes in the values of the physical constants which we had used in order to obtain our UZH06 solution. The largest of these changes concern the values of the charge radii of the 
pion and of the proton. All physical constants will be fixed here to the recommended values of the most recent report of the Particle-Data Group \cite{pdg}.
\item In view of the fact that a large amount of $\pi^\pm p$ elastic-scattering data became available in 2004, one might be under the impression that the important results and conclusions obtained 
with our UZH06 solution (which did not include these data) need to be revised; we will show that this is not the case. Due to numerous problems which we have encountered in the 
analysis of these data \cite{mrw2}, we have decided not to include them in our database.
\item A large amount of CX data appeared after the completion of the UZH06 PSA. In some cases, the corresponding reports of the experimental groups seem to cast doubt on the claims of 
isospin breaking \cite{glk,m}; therefore, it must be investigated whether the newly-obtained measurements invalidate those claims.
\end{itemize}

We must stress that the physical quantities appearing here (i.e., the fit parameters of Sections \ref{sec:K-Matrix_pi+p}-\ref{sec:Model}, the scattering lengths and volumes of Section 
\ref{sec:ModelConstants}, the phase shifts of Section \ref{sec:ModelPhases}, etc.) are not purely-hadronic quantities since they still contain residual em effects, which the stage-I em 
corrections of Refs.~\cite{gmorw} do not remove. These residual effects relate in particular to the fact that, in principle, the (unknown) hadronic masses should be used (in the hadronic 
part of the $\pi N$ interaction) instead of the physical masses of the proton, of the neutron, and of the charged and neutral pion. Unfortunately, it is not possible at the present time 
to assess the significance of these effects. As a result, we must retain the cautious attitude of considering our physical quantities `em-modified hadronic' (as we did in Ref.~\cite{mworg}). 
However, the repetitive use of this term is clumsy. Therefore, we omit it, unless we consider its use necessary as, for instance, in Section \ref{sec:Discussion} and in the captions of 
our tables and figures.

\section{\label{sec:Method}Method}

\subsection{\label{sec:Formalism}Formalism}

The determination of the observables from the hadronic phase shifts has been given in detail in Section 2 of Ref.~\cite{mworg}. For $\pi^+ p$ scattering, one obtains the partial-wave amplitudes 
from Eq.~(1) of that paper and determines the no-spin-flip and spin-flip amplitudes via Eqs.~(2) and (3). Finally, the observables are evaluated from these amplitudes via Eqs.~(13) and (14). For 
$\pi^- p$ elastic scattering, the observables are determined on the basis of Eqs.~(15-20).

\subsection{\label{sec:MF}Minimisation function and scale factors}

Similarly to our previous PSA \cite{mworg}, we make use of the minimisation function given by the Arndt-Roper formula \cite{ar}. The contribution of the $j^{th}$ data set to the overall $\chi^2$ is:
\begin{equation} \label{eq:chijsq}
\chi_j^2=\sum_{i=1}^{N_j} \left\{ \frac{ z_jy_{ij}^{th}-y_{ij}^{exp} }{\delta y_{ij}^{exp} } \right\}^2 + \left( \frac{z_j-1}{\delta z_j} \right)^2 \, ,
\end{equation}
where $y_{ij}^{exp}$ denotes the $i^{th}$ data point of the $j^{th}$ data set, $y_{ij}^{th}$ the corresponding fitted value (also referred to as `theoretical'), $\delta y_{ij}^{exp}$ the statistical 
uncertainty of the $y_{ij}^{exp}$ data point, $z_j$ a scale factor for the relative normalisation applying to the entire data set, $\delta z_j$ the corresponding uncertainty (reported or assigned), 
and $N_j$ the number of data points in the data set. The fitted values $y_{ij}^{th}$ are obtained by means of parameterised forms of the $s$- and $p$-wave amplitudes. The values of the scale factor 
$z_j$ are determined (for each data set separately) in such a way as to minimise $\chi_j^2$. For each data set, there is a unique solution for $z_j$:
\begin{equation} \label{eq:zj}
z_j = \frac{\sum_{i=1}^{N_j} y_{ij}^{th} y_{ij}^{exp} / (\delta y_{ij}^{exp} )^2 + (\delta z_j)^{-2}} {\sum_{i=1}^{N_j} (y_{ij}^{th} / \delta y_{ij}^{exp})^2 + (\delta z_j)^{-2}} \, ,
\end{equation}
which leads to
\begin{equation} \label{eq:chijsqmin}
(\chi_j^2)_{min} = \sum_{i=1}^{N_j} \frac{ (y_{ij}^{th}-y_{ij}^{exp})^2}{(\delta y_{ij}^{exp})^2 } - \frac {\left\{ \sum_{i=1}^{N_j}y_{ij}^{th}(y_{ij}^{th}-y_{ij}^{exp}) / (\delta y_{ij}^{exp} )^2 \right\}^2} 
{ \sum_{i=1}^{N_j}(y_{ij}^{th}/\delta y_{ij}^{exp})^2 + (\delta z_j)^{-2} } \, .
\end{equation}
The overall $\chi^2=\sum_{j=1}^{N} (\chi_j^2)_{min}$ (where $N$ stands for the number of data sets used in the fit) is a function of the parameters entering the modelling of the $s$- 
and $p$-wave amplitudes; these parameters were varied until $\chi^2$ attained its minimal value $\chi^2_{min}$.

The part of $(\chi_j^2)_{min}$ which represents the pure random fluctuations in the measurements of the $j^{th}$ data set (i.e., the `unexplained variation' in standard regression terminology) may be 
obtained from Eq.~(\ref{eq:chijsqmin}) in the limit $\delta z_j \to \infty$, which is equivalent to removing the term $(\delta z_j)^{-2}$ from the denominator of the second term on the right-hand side 
(rhs) of the expression; we will denote this value by $(\chi_j^2)_{st}$. The variation which is contained in $(\chi_j^2)_{min}$ in excess of $(\chi_j^2)_{st}$ must be associated with the contribution 
from the rescaling (floating) of the data set as a whole; the expression for $(\chi_j^2)_{sc} \equiv (\chi_j^2)_{min} - (\chi_j^2)_{st}$ is
\begin{equation} \label{eq:chijsqsc}
(\chi_j^2)_{sc} = \frac{(\delta z_j)^{-2} \left\{ \sum_{i=1}^{N_j}y_{ij}^{th}(y_{ij}^{th}-y_{ij}^{exp}) / (\delta y_{ij}^{exp} )^2 \right\}^2}
{\sum_{i=1}^{N_j}(y_{ij}^{th}/\delta y_{ij}^{exp})^2 \left\{ \sum_{i=1}^{N_j}(y_{ij}^{th}/\delta y_{ij}^{exp})^2 + (\delta z_j)^{-2} \right\} } \, .
\end{equation}
The scale factors which minimise only the first term on the rhs of Eq.~(\ref{eq:chijsq}) are obtained from Eq.~(\ref{eq:zj}) in the limit $\delta z_j \to \infty$:
\begin{equation} \label{eq:zjopt}
\hat{z}_j = \frac{\sum_{i=1}^{N_j} y_{ij}^{th} y_{ij}^{exp} / (\delta y_{ij}^{exp} )^2 } {\sum_{i=1}^{N_j} (y_{ij}^{th} / \delta y_{ij}^{exp})^2 } \, .
\end{equation}

The scale factors of Eq.~(\ref{eq:zj}) are appropriate when investigating the goodness of the overall reproduction of a data set in terms of a reference solution (yielding the fitted values). On the 
contrary, those of Eq.~(\ref{eq:zjopt}) give maximal freedom to the baseline solution when determining the offset of a data set (with respect to that solution) and, as such, are more suitable when the 
emphasis is placed on the absolute normalisation, rather than on the overall reproduction, of a data set.

The statistical uncertainty in the evaluation of the scale factors $\hat{z}_j$ is given by
\begin{equation} \label{eq:zjoptunc}
\delta\hat{z}_j = \frac{1}{\sqrt{\sum_{i=1}^{N_j} (y_{ij}^{th} / \delta y_{ij}^{exp})^2 }} \, .
\end{equation}
When comparing the absolute normalisation of a data set with a reference solution, the statistical uncertainty of Eq.~(\ref{eq:zjoptunc}) must be taken into account, along with the normalisation uncertainty 
$\delta z_j$ of that data set. When investigating the absolute normalisation of some specific data sets, namely of those with suspicious absolute normalisation (e.g., see Section 
\ref{sec:Bertin}), the total uncertainty $\Delta \hat{z}_j = \sqrt{\delta\hat{z}_j^2 + \delta z_j^2}$ will be used.

\subsection{\label{sec:Tests}Statistical tests}

For a data set containing $N_j$ data points, $(N_j+1)$ measurements have actually been made, the additional one relating to the absolute normalisation of the data set. Since the fit involves the fixing 
of each $z_j$ at the value given in Eq.~(\ref{eq:zj}), the proper number of degrees of freedom (hereafter, the acronym DOF will stand for `degree(s) of freedom', whereas NDF for the `number of DOF'), 
associated with the $j^{th}$ data set, is just $N_j$. This implies that the quantity $(\chi_j^2)_{min}$ of Eq.~(\ref{eq:chijsqmin}) is expected to follow the $\chi^2$ distribution with $N_j$ DOF.

The essential difference between the present study and our previous PSA \cite{mworg} is that only one statistical test for each data set will be performed here, the one involving its contribution 
$(\chi_j^2)_{min}$ to the overall $\chi^2$; in Ref.~\cite{mworg}, we instead performed tests for the shape and for the normalisation of each data set.

The p-value which is evaluated on the basis of $(\chi_j^2)_{min}$ and $N_j$ will be compared with the confidence level $\mathrm{p}_{min}$ for the acceptance of the null hypothesis (no 
statistically-significant effects); in case that the extracted p-value is below $\mathrm{p}_{min}$, the DOF with the largest contribution (to the $(\chi_j^2)_{min}$ value) will be eliminated in 
the subsequent fit. Only one point will be removed at each step, and the optimisation will be repeated. Data sets which do not give acceptable p-values (i.e., above $\mathrm{p}_{min}$) after the 
elimination of two of their data points (the absolute normalisation is also considered to be one data point) will be removed from the database.

The second difference to Ref.~\cite{mworg}, as far as the data analysis is concerned, relates to the choice of the confidence level which is assumed in the statistical tests; in Ref.~\cite{mworg}, 
$\mathrm{p}_{min}$ was set to about $2.70 \cdot 10^{-3}$ (which is equivalent to a $3 \sigma$ effect in the normal distribution). Herein, we will instead adopt the $\mathrm{p}_{min}$ value which is 
associated with a $2.5 \sigma$ effect; this value is approximately equal to $1.24 \cdot 10^{-2}$, that is, slightly larger than $1.00 \cdot 10^{-2}$, which is `commonly' (among statisticians) 
associated with the outset of statistical significance. In any case, only a few data (five DOF of the $\pi^\pm p$ elastic-scattering databases) are affected by this more stringent acceptance criterion.

\section{\label{sec:Results}Results}

The repetitive use of the full description of the databases is largely facilitated if we adhere to the following notation: DB$_+$ for the $\pi^+ p$ database; DB$_-$ for the $\pi^- p$ elastic-scattering 
database; DB$_{+/-}$ for the combined $\pi^\pm p$ elastic-scattering databases.

The initial DB$_+$ comprises differential cross sections (DCSs) \cite{ega}-\cite{cj}, analysing powers (APs) \cite{mes,rw}, partial-total cross sections (PTCSs) \cite{bjk,ef}, and total (in fact 
total-nuclear) cross sections (TCSs) \cite{cwbbd,ep}. The initial DB$_+$ consists of $364$ data points, distributed among $54$ data sets, $26$ of which relate to the DCS, $3$ to the AP, and 
$25$ (all one- or two-point data sets) to the PTCS and TCS. The initial DB$_-$ consists of $336$ data points distributed among $36$ data sets, i.e., $27$ for the DCS (\cite{jsf,jtb}, 
\cite{uw}-\cite{cj}, and \cite{mj}) and $9$ for the AP (\cite{mes} and \cite{jca}-\cite{jdp}).

We now list the measurements which have not been included in our analysis.
\begin{itemize}
\item The self-consistency of the $\pi^\pm p$ elastic-scattering DCS measurements of Ref.~\cite{denz} will be addressed in detail elsewhere \cite{mrw2}. This is an enormous piece of experimental data, 
comparable in quantity to the database we established in our UZH06 PSA. Prior to their incorporation into a self-consistent set of data, the self-consistency of these measurements (as well as their 
compatibility with the established database) must be verified; in Ref.~\cite{mrw2}, we will come to a negative result.
\item The $70$ $\pi^+ p$ DCS measurements of Ref.~\cite{brt} are obvious outliers in all exclusive analyses of the low-energy data; when using `traditional' statistics, the inclusion of these 
values in a PSA is bound to introduce spurious effects (e.g., drifting of the parameters during the optimisation, entrapment of the minimisation algorithms in local minima, failing fits, etc.). 
We will further comment on these data in Section \ref{sec:Bertin}.
\item The $6$ ($3$ for $\pi^+ p$ at $94.50$ MeV, $3$ for $\pi^- p$ at $88.50$ MeV) DCS measurements of Ref.~\cite{bussey} have not appeared in a form which would enable their straightforward 
inclusion in our database. Furthermore, we are not convinced that the original DCS data may be retrieved by simply adding the contributions appearing in columns $4$ and $5$ of Table $1$ of 
Ref.~\cite{bussey}.
\item Given that the $9$ existing $\pi^- p$ PTCSs and TCSs contain a component from CX scattering, they cannot be used; the inclusion of these data in any part of the analysis would 
perplex the discussion on the violation of the isospin invariance in the hadronic part of the $\pi N$ interaction \cite{mrw3}.
\item The inclusion of the $28$ AP measurements of Meier \etal~(MEIER04) \cite{meier} in the fits would necessitate substantial modifications in the database structure and in the analysis software. 
This is due to two reasons: a) each of the three data sets, to which the measurements of Ref.~\cite{meier} must be assigned, contains data taken at more than one beam energy and b) the last of the data 
sets contains measurements for both elastic-scattering reactions. Given the difficulty at present to include these measurements in our fits, we can only use them in testing the overall consistency 
of our approach across energies. (If a significant amount of $\pi^- p$ measurements had been acquired in that experiment, we could have investigated the consistency of our approach across reactions.) 
We will return to this subject in Section \ref{sec:Meier}.
\item We will also not use in the fits the scattering length obtained from the experimental result for the strong shift of the $1s$ state in pionic hydrogen \cite{ss}, after it has been corrected in 
Ref.~\cite{orwmg} also taking into account the proper contributions of the $\gamma n$ channel; the difference to the $\tilde{a}^{cc}$ value of our UZH06 PSA (which is almost identical to the value 
we will obtain in this work) has been addressed in Ref.~\cite{orwmg}.
\end{itemize}

In order to give the data maximal freedom in the process of identifying the outliers, the two elastic-scattering reactions will be analysed separately using simple parameterisations of the $s$- and $p$-wave 
$K$-matrix elements. The (small at low energies) $d$ and $f$ waves have been fixed herein from the current SAID solution (WI08) \cite{abws}. In the SAID analysis, the energy dependence of the $d$ and 
$f$ phase shifts is determined from the region $T > 100$ MeV (i.e., from energies where these contributions are sizable). The largest of these phase shifts in the energy interval of this analysis, D15, 
does not exceed $0.27^\circ$.

For the purpose of fitting, the standard MINUIT package \cite{jms} of the CERN library was used (FORTRAN version). Each optimisation was achieved on the basis of the (robust) SIMPLEX-MINIMIZE-MIGRAD-MINOS chain. 
All fits of the present work terminated successfully.

\subsection{\label{sec:K-Matrix_pi+p}Fits to the DB$_+$ using the $K$-matrix parameterisations}

The parameterisation, which we will now describe, was introduced (and successfully applied to $\pi^+ p$ scattering) in Ref.~\cite{fm}. For $\pi^+ p$ elastic scattering, the 
$s$-wave phase shift is parameterised as
\begin{equation} \label{eq:delta0+3/2}
q_c \cot \tilde{\delta}_{0+}^{3/2}=(\tilde{a}_{0+}^{3/2})^{-1} + b_3 \epsilon + c_3 \epsilon^2 \, ,
\end{equation}
where $q_c$ and $\epsilon$ are respectively the momentum and the pion kinetic energy in the centre-of-mass (CM) system. The $p_{1/2}$-wave phase shift is parameterised 
according to the form
\begin{equation} \label{eq:delta1-3/2}
\tan \tilde{\delta}_{1-}^{3/2}/q_c = d_{31} \epsilon + e_{31} \epsilon^2 \, .
\end{equation}

Since the $p_{3/2}$ wave contains the $\Delta$($1232$) resonance, a resonant piece in Breit-Wigner form is added to the background term, thus leading to the equation
\begin{equation} \label{eq:delta1+3/2}
\tan \tilde{\delta}_{1+}^{3/2}/q_c = d_{33} \epsilon + e_{33} \epsilon^2 + \frac{\Gamma_{\Delta}m_{\Delta}^2}{q_{\Delta}^3W} \frac{q_c^2}{m_{\Delta}^2-W^2} \, ,
\end{equation}
with $m_{\Delta}=1232$ MeV and $\Gamma_{\Delta}=118$ MeV \cite{pdg}; $q_\Delta$ is the value of the CM momentum at the resonance position. The third term on the rhs of Eq.~(\ref{eq:delta1+3/2}) is the 
standard resonance contribution (e.g., see Ref.~\cite{ew}, p.~31).

The use of the parametric forms in Eqs.~(\ref{eq:delta0+3/2})-(\ref{eq:delta1+3/2}), which do not impose any theoretical constraints (save for the expected low-energy behaviour of the $K$-matrix elements 
and for the Breit-Wigner form used to describe the $\Delta$($1232$)-resonance contribution), ensures that any outliers in the fits cannot be attributed to the inability of these forms to account for the 
energy dependence of the phase shifts; they are instead indicative of experimental problems.

Following the procedure of Section \ref{sec:Tests}, we found that, for the data sets of BRACK90 \cite{jtbbb} at $66.80$ MeV and JORAM95 \cite{cj} at $32.70$ MeV, the p-values were still below $\mathrm{p}_{min}$ 
after the removal of two data points (from each data set); as a result, these two data sets (with eleven and seven data points, respectively) were removed from the database. These two data sets have extremely 
low p-values and clearly stand out from the rest of the $\pi^+ p$ data. Two data sets had to be freely floated, the BRACK86 \cite{jtb} measurements at $66.80$ and $86.80$ MeV. Finally, four additional single data points were removed.

After the elimination of $24$ DOF of the initial database, we obtain a truncated DB$_+$ comprising $52$ data sets with $340$ DOF. The surviving data sets and the corresponding NDF are 
listed in Table \ref{tab:DBpi+p}; the numbers have been taken from the final fit to the truncated DB$_+$ using the $K$-matrix parameterisations.

Since seven parameters were used in order to generate the fitted values, the NDF in the fit to the initial database was $357$; the minimal value of $\chi^2$ was $675.7$. 
For the truncated database with $333$ DOF in the fit, the minimal value of $\chi^2$ was $427.2$. Therefore, the elimination of only $24$ DOF of the initial database leads to an 
impressive decrease of the $\chi^2$ by $248.5$ units. At the same time, the p-value of the fit increased by over $17$ orders of magnitude.

\subsection{\label{sec:K-Matrix_pi-p}Fits to the DB$_-$ using the $K$-matrix parameterisations}

The $I=3/2$ amplitudes, obtained in the final fit to the truncated DB$_+$ using the $K$-matrix parameterisations, were imported into the analysis of the DB$_-$. In this part, another seven parameters were introduced, 
to parameterise the $I=1/2$ amplitudes. As in the $\pi^+ p$ case, these parameters were varied pursuing the minimisation of the $\chi^2$ function. Similar parametric forms were used 
as those given in Eqs.~(\ref{eq:delta0+3/2})-(\ref{eq:delta1+3/2}), with the parameters $\tilde{a}_{0+}^{1/2}$, $b_1$, $c_1$, $d_{13}$, $e_{13}$, $d_{11}$, and $e_{11}$. Of course, 
there is no resonance term in the expression for $\tilde{\delta}_{1+}^{1/2}$; instead, it is necessary to include the contribution of the Roper resonance in $\tilde{\delta}_{1-}^{1/2}$:
\begin{equation} \label{eq:delta1-1/2}
\tan \tilde{\delta}_{1-}^{1/2}/q_c = d_{11} \epsilon + e_{11} \epsilon^2 + \frac{\Gamma_{N}m_{N}^2}{q_{N}^3W} \frac{q_c^2}{m_{N}^2-W^2} \, ,
\end{equation}
with $m_N=1440$ MeV and $\Gamma_N=195$ MeV \cite{pdg}; $q_N$ denotes the CM momentum at the Roper-resonance position. As we are dealing with energies below the pion-production threshold, 
$\Gamma_N$ is the elastic width.

The DB$_-$ was also subjected to the tests described in Section \ref{sec:Tests}. As in the UZH06 PSA, the BRACK90 \cite{jtbbb} $66.80$ MeV data set (containing five data points) was 
marked as an outlier. Two additional single data points had to be removed. Furthermore, the WIEDNER89 \cite{uw} data set had to be freely floated. After the elimination of $8$ DOF of the 
initial DB$_-$, we obtain a truncated database comprising $35$ data sets with $328$ DOF (see Table \ref{tab:DBpi-p}). Since seven parameters were used in order to generate the fitted 
values, the NDF in the fit to the initial database was $329$; the minimal value of $\chi^2$ was $528.1$. For the truncated database with $321$ DOF in the fit, the minimal value of $\chi^2$ 
was $371.0$. At the same time, the p-value of the fit increased by $9$ orders of magnitude.

\subsection{\label{sec:K-Matrix}Common fit to the DB$_{+/-}$ using the $K$-matrix parameterisations}

Judged solely on the basis of the p-values, it appears that the truncated DB$_{+}$ and DB$_{-}$ are not of the same quality. However, there is an appropriate 
statistical measure for such a comparison of two quantities following the $\chi^2$ distribution. In order to prove that the two databases are of different quality (that is, that they have 
not been sampled from the same distribution), the ratio
\begin{equation} \label{eq:grossF}
F=\frac{\chi_+^2/N\!D\!F\!_+}{\chi_-^2/N\!D\!F\!_-}
\end{equation}
must be significantly different from $1$. In this formula, the subscripts `$+$' and `$-$' denote the two elastic-scattering reactions. The ratio $F$ follows Fisher's ($F$) distribution. From the two 
final fits to the truncated databases using the $K$-matrix parameterisations, we obtain the score value of $1.110$ for $N\!D\!F\!_+=333$ and $N\!D\!F\!_-=321$ DOF in the two separate fits, which is translated 
into the p-value of $1.7 \cdot 10^{-1}$. Therefore, the claim about the dissimilarity of the two databases cannot be sustained. As a result, it makes sense to analyse the two 
reactions in terms of a common optimisation scheme.

In order to give the two elastic-scattering reactions equal weight, we multiplied $(\chi^2_j)_{min}$ for each $\pi^+ p$ data set by
\[
w_+=\frac{N\!_+ + N\!_-}{2N\!_+}
\]
and for each $\pi^- p$ elastic-scattering data set by
\[
w_-=\frac{N\!_+ + N\!_-}{2N\!_-} \, ,
\]
where $N\!_+$ and $N\!_-$ represent the NDF in the two databases; we then added these quantities for all the data sets to obtain the overall $\chi^2$ value. The application of these `global' 
weights for the two reactions was made as a matter of principle; given that the values of $N\!_+$ and $N\!_-$ are very close, the effect of this weighting on our results is very small.

The common fit to the truncated DB$_{+/-}$ (detailed in Tables \ref{tab:DBpi+p} and \ref{tab:DBpi-p}) was made, using the $14$ parameters of the $K$-matrix parameterisations given in Sections \ref{sec:K-Matrix_pi+p} 
and \ref{sec:K-Matrix_pi-p}. We did not find any additional data points (or data sets) which had to be removed; we concluded that the truncated DB$_{+/-}$ are self-consistent. The common fit to the data 
yielded a $\chi^2$ value of $792.4$ for $654$ DOF. The set of the excluded DOF represents $4.57 \%$ of the initial database. In the following, we will use this truncated DB$_{+/-}$. 

\subsection{\label{sec:Model}Common fits to the truncated DB$_{+/-}$ using the ETH model}

The modelling of the hadronic part of the $\pi N$ interaction on the basis of the $K$-matrix parameterisations of Sections \ref{sec:K-Matrix_pi+p} and \ref{sec:K-Matrix_pi-p} is suitable as a first test of the 
self-consistency of the two elastic-scattering databases and as an efficient method for the identification of the outliers, yet neither does it provide insight into the underlying physical processes nor can 
it easily incorporate the important theoretical constraint of crossing symmetry. In order to accomplish these two tasks, we will next involve in the analysis a model based on Feynman diagrams, namely the 
ETH model. This model was introduced in Ref.~\cite{glmbg} and was developed further throughout the early 1990s. The ability of the ETH model to account for the low-energy $\pi^\pm p$ elastic-scattering data 
has been convincingly demonstrated over the past two decades.

The main diagrams on which this isospin-invariant model is based are graphs with scalar-isoscalar ($I=J=0$) and vector-isovector ($I=J=1$) $t$-channel exchanges, as well as the $N$ and $\Delta$ $s$- and 
$u$-channel graphs. The main contributions to the partial-wave amplitudes from these diagrams have been given in detail in Ref.~\cite{glmbg}. The small contributions from the six well-established four-star 
$s$ and $p$ higher baryon resonances with masses up to $2$ GeV have also been analytically included in the model; in fact, the only significant contributions come from the Roper resonance. The tensor component 
of the $I=J=0$ $t$-channel exchange was added in Ref.~\cite{m}; after this (insignificant) modification, no changes have been made to the model.

The $I=J=0$ $t$-channel contribution to the amplitudes is approximated in the model by a broad $\pi \pi$ resonance, characterised by two parameters, $G_\sigma$ and $K_\sigma$. Its exact position has practically 
no effect on the description of the $\pi N$ scattering data or on the fitted values of $G_\sigma$ and $K_\sigma$; for a long time, it has been fixed at $860$ MeV. The $I=J=1$ $t$-channel contribution is described 
by the $\rho$-meson, with $m_\rho=775.49$ MeV \cite{pdg}; this contribution introduces two additional parameters: $G_\rho$ and $K_\rho$. The contributions of the $s$- and $u$-channel graphs with an intermediate 
$N$ involve the $\pi N N$ coupling constant $g_{\pi N N}$ and one additional parameter $x$ representing the pseudoscalar admixture in the $\pi N N$ vertex; for pure pseudovector coupling, $x=0$. Finally, the 
contributions of the graphs with an intermediate $\Delta$ state introduce the coupling constant $g_{\pi N \Delta}$ and the parameter $Z$ (which is associated with the spin-$1/2$ admixture in the $\Delta$ field). 
The higher baryon resonances do not introduce any parameters.

When a common fit of the ETH model to the data is made using all eight aforementioned parameters, it turns out that there is a strong correlation between $G_\sigma$, $G_\rho$, and $x$; due to this correlation, it is not possible 
to determine the values of all three quantities. We have chosen to set $x$ to $0$; this choice is usually adopted in effective-field theoretical models of low-energy $\pi N$ scattering. The common fits of 
the ETH model to the truncated DB$_{+/-}$ will be performed on the basis of seven parameters: $G_\sigma$, $K_\sigma$, $G_\rho$, $K_\rho$, $g_{\pi NN}$, $g_{\pi N \Delta}$, and $Z$.

\subsubsection{\label{sec:ModelParameters}Model parameters}

The choice of the probability value below which data points must be excluded is difficult. We have adopted here the value of $\mathrm{p}_{min}$ corresponding to a $2.5 \sigma$ effect in the normal distribution. 
Recognising the subjectiveness in this choice, we consider that, in order to have confidence in the reliability of our analysis, it is necessary to verify that the fitted values of the seven model parameters 
remain stable over a reasonably broad range of $\mathrm{p}_{min}$ values; we followed the same strategy in Ref.~\cite{mworg}. Thus, in addition to $\mathrm{p}_{min} \approx 1.24 \cdot 10^{-2}$, the analysis was 
performed with a database reduced by using the $\mathrm{p}_{min}$ values of about $2.70 \cdot 10^{-3}$ (equivalent to a $3 \sigma$ effect in the normal distribution) and $4.55 \cdot 10^{-2}$ (equivalent 
to a $2 \sigma$ effect in the normal distribution)~\footnote{It must be mentioned that, in Ref.~\cite{mworg}, the results for the different $\mathrm{p}_{min}$ levels had been obtained by applying cuts to the 
distribution of the residuals, as this distribution came out in the $\mathrm{p}_{min} \approx 2.70 \cdot 10^{-3}$ solution. On the contrary, the analysis here is performed separately for the different $\mathrm{p}_{min}$ levels.}.

Table \ref{tab:ModelParameters} shows the values of the seven model parameters for the common fits to the truncated DB$_{+/-}$ for the three selected values of $\mathrm{p}_{min}$. The uncertainties shown 
correspond to $\mathrm{p}_{min} \approx 1.24 \cdot 10^{-2}$. In fact, when the uncertainties are calculated with the Birge factor $\sqrt{\chi^2/\mathrm{NDF}}$ included, they do not vary much with the value 
of $\mathrm{p}_{min}$. As $\mathrm{p}_{min}$ is increased, the truncated database fitted shrinks and so the raw uncertainties increase. However, the factor $\sqrt{\chi^2/\mathrm{NDF}}$ decreases as the fit 
quality improves (despite the decrease in ${\mathrm{NDF}}$) and the two effects largely compensate. Table \ref{tab:ModelParameters} shows that the results of the fit are reasonably stable as the criterion for the 
rejection of data points is varied~\footnote{The effects, which are seen when increasing the $\mathrm{p}_{min}$ value from the equivalent of a $2.5$ to a $3 \sigma$ effect in the normal distribution, are 
due to the removal of the large JORAM95 $\pi^+ p$ data set at $44.60$ MeV in the analysis of the data using the $K$-matrix parameterisations of Section \ref{sec:K-Matrix_pi+p}. A third data point from this data set 
(the measurement at $14.26^\circ$) must be excluded at the last step of the iteration, thus resulting in the removal of the entire data set; were this data set not excluded, there would have been 
almost no change in the parameter values shown in Table \ref{tab:ModelParameters}.}.

We see from Table \ref{tab:ModelParameters} that the values of $g_{\pi NN}$ are compatible; converted to the usual pseudovector coupling constant~\footnote{Some authors redefine $f_{\pi NN}^2$, absorbing in 
it the factor $4 \pi$.}, our result for $\mathrm{p}_{min} \approx 1.24 \cdot 10^{-2}$ is
\[
\frac{f_{\pi N N}^2}{4 \pi}= \left( \frac{\mu_c}{2 m_p} \right)^2 \frac{g_{\pi NN}^2}{4 \pi} = 0.0726(14) \, ,
\]
where $\mu_c$ denotes the mass of the charged pion and $m_p$ that of the proton. This result agrees well with the value we had obtained in our previous PSA \cite{mworg}.

The correlation (Hessian) matrix for the seven parameters of the ETH model is given in Table \ref{tab:HessianMatrix}; the numbers correspond to the fit for $\mathrm{p}_{min} \approx 1.24 \cdot 10^{-2}$. 
This matrix, together with the uncertainties given in Table \ref{tab:ModelParameters}, enables the determination of predictions (and of their associated uncertainties) for the threshold constants, for the 
hadronic phase shifts and amplitudes, and for the observables at any combination of the energy and of the scattering angle. Table \ref{tab:ModelParameters} shows that the value of $K_\sigma$ is consistent 
with $0$; the quality of the fit would deteriorate very little if this parameter were set to $0$. The value of $G_\sigma$ is very little correlated with the values of the other five parameters. However, 
these parameters ($G_\rho$, $K_\rho$, $g_{\pi NN}$, $g_{\pi N \Delta}$, and $Z$) are all strongly correlated with each other. (As expected, the correlations among the model parameters are smaller when no 
floating of data sets is allowed in the fit.)

Our results for the seven model parameters have shown remarkable stability over the years, from the period when the fits were performed to old, outdated phase shifts (e.g., those of Refs.~\cite{kp}) to 
the present times when the low-energy $\pi N$ measurements are directly fitted to. The database itself has changed significantly over the past twenty or so years, with the important contributions from 
experiments performed at the meson factories by different research groups, which made use of different apparatuses and techniques in their experiments. Our method of applying the em corrections has also 
changed. Finally, various approaches have been implemented in the optimisation (e.g., in the choice of the minimisation function, from `standard' $\chi^2$ functions \cite{fm}, to robust statistics without 
any rejection of data \cite{m}, to the use of the Arndt-Roper formula \cite{ar} along with pruning the $\pi N$ databases \cite{mworg}). The observed stability is indicative of the robustness one obtains in 
the results when involving the ETH model in the analysis of the low-energy $\pi N$ data.

\subsubsection{\label{sec:ModelConstants}Threshold constants}

From the parameters of the ETH model and their uncertainties given in Table \ref{tab:ModelParameters} for $\mathrm{p}_{min} \approx 1.24 \cdot 10^{-2}$, as well as the correlation matrix given in Table 
\ref{tab:HessianMatrix}, we determined the isoscalar and isovector $s$-wave scattering lengths and the isoscalar(isovector)-scalar(vector) $p$-wave scattering volumes. The results are:
\[
\frac{1}{3}\:\tilde{a}_{0+}^{1/2}+\frac{2}{3}\:\tilde{a}_{0+}^{3/2}=0.0033(12)\: \mu_c^{-1} ,
\]
\[
-\frac{1}{3}\:\tilde{a}_{0+}^{1/2}+\frac{1}{3}\:\tilde{a}_{0+}^{3/2}=-0.07698(60)\: \mu_c^{-1} , \]
\begin{equation}
\frac{1}{3}\:\tilde{a}_{1-}^{1/2}+\frac{2}{3}\:\tilde{a}_{1-}^{3/2}+\frac{2}{3}\:\tilde{a}_{1+}^{1/2}+\frac{4}{3}\:\tilde{a}_{1+}^{3/2}=0.2039(19)\: \mu_c^{-3} ,
\label{eq:atildas}
\end{equation}
\[
-\frac{1}{3}\:\tilde{a}_{1-}^{1/2}+\frac{1}{3}\:\tilde{a}_{1-}^{3/2}-\frac{2}{3}\:\tilde{a}_{1+}^{1/2}+\frac{2}{3}\:\tilde{a}_{1+}^{3/2}=0.1728(18)\: \mu_c^{-3} ,
\]
\[
\frac{1}{3}\:\tilde{a}_{1-}^{1/2}+\frac{2}{3}\:\tilde{a}_{1-}^{3/2}-\frac{1}{3}\:\tilde{a}_{1+}^{1/2}-\frac{2}{3}\:\tilde{a}_{1+}^{3/2}=-0.1830(19)\: \mu_c^{-3} ,
\]
\[
-\frac{1}{3}\:\tilde{a}_{1-}^{1/2}+\frac{1}{3}\:\tilde{a}_{1-}^{3/2}+\frac{1}{3}\:\tilde{a}_{1+}^{1/2}-\frac{1}{3}\:\tilde{a}_{1+}^{3/2}=-0.06724(83)\: \mu_c^{-3} .
\]

Converting these results to the familiar spin-isospin quantities, we obtain
\[
\tilde{a}_{0+}^{3/2}=-0.0737(16)\: \mu_c^{-1} ,\qquad \tilde{a}_{0+}^{1/2}=0.1573(13)\: \mu_c^{-1} ,
\]
\begin{equation}
\tilde{a}_{1+}^{3/2}=0.2090(20)\: \mu_c^{-3} ,\qquad \tilde{a}_{1+}^{1/2}=-0.03110(66)\: \mu_c^{-3} ,
\label{eq:atildasvalues}
\end{equation}
\[
\tilde{a}_{1-}^{3/2}=-0.04124(78)\: \mu_c^{-3} ,\qquad \tilde{a}_{1-}^{1/2}=-0.0796(16)\: \mu_c^{-3} .
\]

Our results for the $s$-wave scattering lengths $\tilde{a}_{0+}^{3/2}$ and $\tilde{a}_{0+}^{1/2}$ in Eqs.~(\ref{eq:atildasvalues}) are compatible with those obtained in Refs.~\cite{fm,m,mworg}; these 
values have been very stable over the last fifteen years. The large quantity of the elastic-scattering data below $100$ MeV obtained at pion factories since 1980, when analysed without influences from 
the data obtained at higher energies, leads to results for the $s$-wave scattering lengths (and hadronic phase shifts) which are significantly different from those extracted via dispersion relations 
after also including the $\pi^- p$ charge-exchange database in the analysis and using the measurements up to the few-GeV region.

From the results in Eqs.~(\ref{eq:atildasvalues}), we obtain
\[
\tilde{a}^{cc} = \frac{2}{3} \, \tilde{a}_{0+}^{1/2} + \frac{1}{3} \, \tilde{a}_{0+}^{3/2} = 0.0803(11)\: \mu_c^{-1} \, ,
\]
in good agreement with the value extracted in Ref.~\cite{mworg}. We have already discussed \cite{mworg,orwmg} the general disagreement (which, at present, is equivalent to a $4.6 \sigma$ effect in the 
normal distribution) of our $\tilde{a}^{cc}$ value, extracted from $\pi^\pm p$ elastic-scattering measurements, with the result obtained in the pionic-hydrogen experiments at the Paul Scherrer Institut 
(PSI) \cite{ss}, after the corrections of Ref.~\cite{orwmg} are applied. We comment further on this issue in Section 6.1 of Ref.~\cite{mrw3}.

\subsubsection{\label{sec:ModelPhases}Hadronic phase shifts}

The $s$- and $p$-wave hadronic phase shifts, obtained from the common fit to the truncated DB$_{+/-}$ using the ETH model, are given in Table \ref{tab:PhaseShifts}. These phase shifts are also shown in 
Figs.~\ref{fig:a}-\ref{fig:f}, together with the current SAID solution (WI08) \cite{abws} and their five single-energy values (whenever available).

It is evident from Figs.~\ref{fig:a} and \ref{fig:b} that our values of the $s$-wave hadronic phase shifts $\tilde{\delta}_{0+}^{3/2}$ and $\tilde{\delta}_{0+}^{1/2}$ differ significantly from the SAID 
results. Our values of $\tilde{\delta}_{0+}^{3/2}$ are less negative, but converge towards the SAID values as the energy approaches $100$ MeV; for $\tilde{\delta}_{0+}^{1/2}$, our values are consistently smaller.

For the $p$-wave hadronic phase shifts $\tilde{\delta}_{1+}^{3/2}$, $\tilde{\delta}_{1-}^{3/2}$, and $\tilde{\delta}_{1+}^{1/2}$, inspection of Figs.~\ref{fig:c}-\ref{fig:e} shows that 
there is general agreement between the two solutions (the differences do not exceed about $0.1^\circ$). The significant difference in the $p$-wave part of the interaction occurs for $\tilde{\delta}_{1-}^{1/2}$; 
our values are more negative. The values of $\tilde{\delta}_{1-}^{1/2}$, obtained in UZH06 and ZUAS12, are slightly different. This change is almost entirely due to the current use of a lower $\Gamma_N$ 
value; at the time when the UZH06 PSA was carried out, the Particle-Data Group recommended the $\Gamma_N$ value of $227.5$ MeV.

Our phase-shift values are expected to be of interest in analyses involving the low-energy $\pi N$ interaction, as well as in the determination of the $\pi N$ $\sigma$ term (e.g., see Ref.~\cite{aco}).

We will now express our criticism concerning the SAID results at low energies.
\begin{itemize}
\item One has the impression that new $\pi N$ measurements enter the SAID database without regard of whether they comprise a self-consistent set and/or of whether they are at least marginally-compatible 
with the data which are already part of their database. Such a strategy could be less problematic, if they had implemented robust statistics in their data analysis; instead, their results are obtained 
with a `standard' $\chi^2$ function (i.e., the Arndt-Roper formula), and are thus expected to be sensitive to the presence of outliers in the database (in particular, to the presence of \emph{one-sided} outliers).
\item The SAID results at low energies are literally swamped by the measurements at higher energies. In case that the floating of the data sets is allowed (as is when using the Arndt-Roper formula in the optimisation), 
it is unavoidable that the low-energy behaviour of the $\pi N$ amplitudes will be influenced from higher energies. As a result, the low-energy experiments 
will be scaled systematically in such a way as to match the trends of the amplitudes suitable for the higher-energy data.
\item We have not found a published plot from the SAID group, showing the energy dependence of their scale factors below $100$ MeV.
\item The distribution of the normalised residuals and the function which is used in the optimisation are intimately connected. In case that a $\chi^2$ minimisation function is used, the distribution of the 
normalised residuals must, in a self-consistent analysis, be a Gaussian centred at $0$. Any effects observed in the distribution of the normalised residuals (e.g., significant offset, asymmetry of the distribution) 
are indicative of problems during the fitting procedure. We have not found this information in the SAID reports or in their web site.
\item By forcing the data from all three reactions into an isospin-invariant analysis scheme, SAID cannot explore a possible violation of the isospin invariance in the hadronic part of the $\pi N$ interaction.
\end{itemize}

\subsubsection{\label{sec:ScaleFactors}Scale factors and normalised residuals}

We will now comment on the distribution of the scale factors $z_j$ obtained from the common fits using the ETH model. In a `healthy' fit made on the basis of the Arndt-Roper formula, the data sets which must be 
scaled `upwards' should (more or less) be balanced by those which must be scaled `downwards'. Additionally, the energy dependence of the scale factors over the energy range of the analysis should not be significant. 
If these prerequisites are not fulfilled, the parametric forms used in the fits cannot adequately reproduce the data over the entire energy range. For both the $\pi^+ p$ (Fig.~\ref{fig:sfpip}) and $\pi^- p$ 
(Fig.~\ref{fig:sfpim}) elastic-scattering data sets, the values of $z_j$ which lie above and below $1$ roughly balance each other and there is no discernible energy dependence~\footnote{The linear fit to the scale 
factors for the $\pi^+ p$ reaction yields the intercept of $1.012 \pm 0.018$ and the slope of $(-0.9 \pm 2.4) \cdot 10^{-4}$ MeV$^{-1}$; the linear fit to the scale factors for the $\pi^- p$ elastic-scattering reaction 
yields the intercept of $1.031 \pm 0.018$ and the slope of $(-4.5 \pm 2.4) \cdot 10^{-4}$ MeV$^{-1}$.}. Evidently, there is no subrange of the entire $30$ to $100$ MeV energy interval in which the data is better or 
worse fitted than for the rest of the range.

A second issue which must be investigated in a fit involving the minimisation of any $\chi^2$ function is whether the distribution of the normalised residuals $r_{ij}$, defined as
\[
r_{ij} = \frac{ z_jy_{ij}^{th}-y_{ij}^{exp} }{\delta y_{ij}^{exp} }
\]
(see Eq.~(\ref{eq:chijsq})), evaluated on the basis of the optimal parameter values, is Gaussian~\footnote{Of course, the values of $(z_j-1)/\delta z_j$, appearing in the second term on the rhs of Eq.~(\ref{eq:chijsq}), 
must also be included in the distribution of the normalised residuals.}. In practice, one fits a Gaussian function to the distribution of the normalised residuals, i.e., 
\[
f(r) = A e^{-B(r-\bar{r})^2} \, ,
\]
and investigates the quality of the fit (expressed through the corresponding $\chi^2$ value and the NDF in the fit), as well as the asymmetry of the fitted distribution (e.g., expressed through the 
deviation of the extracted value of $\bar{r}$ from $0$). The distribution of the normalised residuals is shown in Fig.~\ref{fig:residuals}, along with the optimal Gaussian function. The $\chi^2$ value of 
this fit was $33.2$ for $22$ DOF in the fit, whereas $\bar{r}=(1.9 \pm 4.7) \cdot 10^{-2}$, i.e., compatible with $0$. For the sake of completeness, we also give the optimal value and the uncertainty of the 
parameter $B$: $B=0.544 \pm 0.038$; the expectation value for $B$ is $0.5$.

\subsubsection{\label{sec:Meier}Reproduction of the MEIER04 measurements}

We will now discuss the MEIER04 measurements (which have not been included in our fits). We have created Monte-Carlo predictions for the AP corresponding to each of their $28$ data points. For the three 
experimental data sets, the resulting values of $\chi^2_{min}$ were $12.5$, $7.0$, and $16.1$, for $12$, $6$, and $10$ DOF, respectively. The values of the scale factor for the three data sets (in the same order) 
are $1.011$, $0.973$, and $1.039$; the reported normalisation uncertainty of the data is $3.5 \%$. It is evident that our hadronic phase shifts reproduce the MEIER04 measurements very well; even the smallest 
p-value does not fall below $9.6 \cdot 10^{-2}$. This is a good test of the consistency of our approach in describing the energy and angular dependence of the $\pi^\pm p$ elastic-scattering data.

\subsubsection{\label{sec:Bertin}Reproduction of the BERTIN76 measurements}

We will now comment on the reproduction of the old measurements of Bertin \etal~(BERTIN76) \cite{brt}, which are `traditionally' considered outliers in almost all modern PSAs. Reported in Ref.~\cite{brt} were $\pi^+ p$ 
DCS measurements obtained in a broad angular interval, at seven beam energies ($20.80$, $30.50$, $39.50$, $51.50$, $67.40$, $81.70$, and $95.90$ MeV). The experimental group did not report any normalisation uncertainties 
for their measurements; it cannot be excluded that, at those times, the absolute normalisation was not seriously investigated in the $\pi N$ experiments.

We assigned rough normalisation uncertainties to the BERTIN76 data, on the basis of the values obtained from the modern low-energy experiments which properly reported this quantity, and analysed these measurements 
as if they comprised the entire DB$_+$ at low energies. The analysis in terms of our general $K$-matrix parameterisations of Section \ref{sec:K-Matrix_pi+p} revealed that the data set at $67.40$ MeV had to 
be eliminated due to its (very bad) shape. We were able to fit the remaining data successfully, and obtained the final $\chi^2$ value of $55.0$ for $53$ DOF in the fit.

We subsequently investigated how well the phase-shift solution of the present study reproduces the BERTIN76 data; it failed. The basic problem with the BERTIN76 measurements lies with their absolute normalisation, 
not with their shape (though the shapes of the data sets at $39.50$ and at $95.90$ MeV do not pass our $\mathrm{p}_{min}$ criterion). For instance, for the data set at $20.80$ MeV, $\hat{z}_j=1.349$ when the overall 
uncertainty $\Delta \hat{z}_j$ (defined at the end of Section \ref{sec:MF}) is equal to $0.081$; this is the most striking discrepancy in the data. The extracted $\hat{z}_j$ factors decrease almost linearly with $T$, 
reaching $\hat{z}_j=1.027$ at $95.90$ MeV. In view of these large energy-dependent effects in their absolute normalisation, we will continue excluding the BERTIN76 data in our PSAs.

\section{\label{sec:Discussion}Discussion}

In the present work, comprising the first of three papers dealing with issues of the pion-nucleon ($\pi N$) interaction below pion laboratory kinetic energy of $100$ MeV, we report the results of a new phase-shift 
analysis (PSA) of the $\pi^\pm p$ elastic-scattering data, using the electromagnetic (em) corrections of Refs.~\cite{gmorw}.

There are two main differences to the approach we set forth in our previous PSA \cite{mworg}, both pertaining to the method for the exclusion of outliers, single data points and entire data sets, in the optimisation phase. 
We now perform only \emph{one} test for each data set on the only relevant quantity, namely on its contribution $(\chi_j^2)_{min}$ to the overall $\chi^2$. In Ref.~\cite{mworg}, we instead performed several tests (on shape, 
normalisation, etc.) for each data set; the use of only one test led to the exclusion of fewer data compared to Ref.~\cite{mworg}. The second difference concerns the imposition of a more stringent acceptance criterion of 
the null hypothesis in the statistical tests. Herein, we raised the minimal p-value ($\mathrm{p}_{min}$) for the acceptance of the null hypothesis from the equivalent of a $3$ to a $2.5 \sigma$ effect in the normal distribution; 
the latter value is closer to the `common' choice (of most statisticians) as the outset of statistical significance.

Similarly to Ref.~\cite{mworg}, we first investigated the self-consistency of the low-energy $\pi^\pm p$ elastic-scattering databases, via two separate analyses carried out (first) on the $\pi^+ p$ and (subsequently) on the 
$\pi^- p$ elastic-scattering data using simple $K$-matrix parameterisations. We found that it is possible to obtain self-consistent databases by removing the measurements of two $\pi^+ p$ and one $\pi^- p$ data sets, as well as a few 
single data points; the removal of these outliers resulted in enormous reductions in the minimal $\chi^2$ values for the separate fits to the two elastic-scattering databases using the $K$-matrix parameterisations. We give all the details 
concerning the accepted data sets in Tables \ref{tab:DBpi+p} and \ref{tab:DBpi-p}; these details may be useful in other analyses. The aforementioned results were obtained without imposing any theoretical constraints, save for the 
expected low-energy behaviour of the $s$- and $p$-wave $K$-matrix elements and the Breit-Wigner form (see Eqs.~(\ref{eq:delta1+3/2}) and (\ref{eq:delta1-1/2})) for the contributions of the resonant terms.

\begin{figure}
\begin{center}
\includegraphics [width=15.5cm] {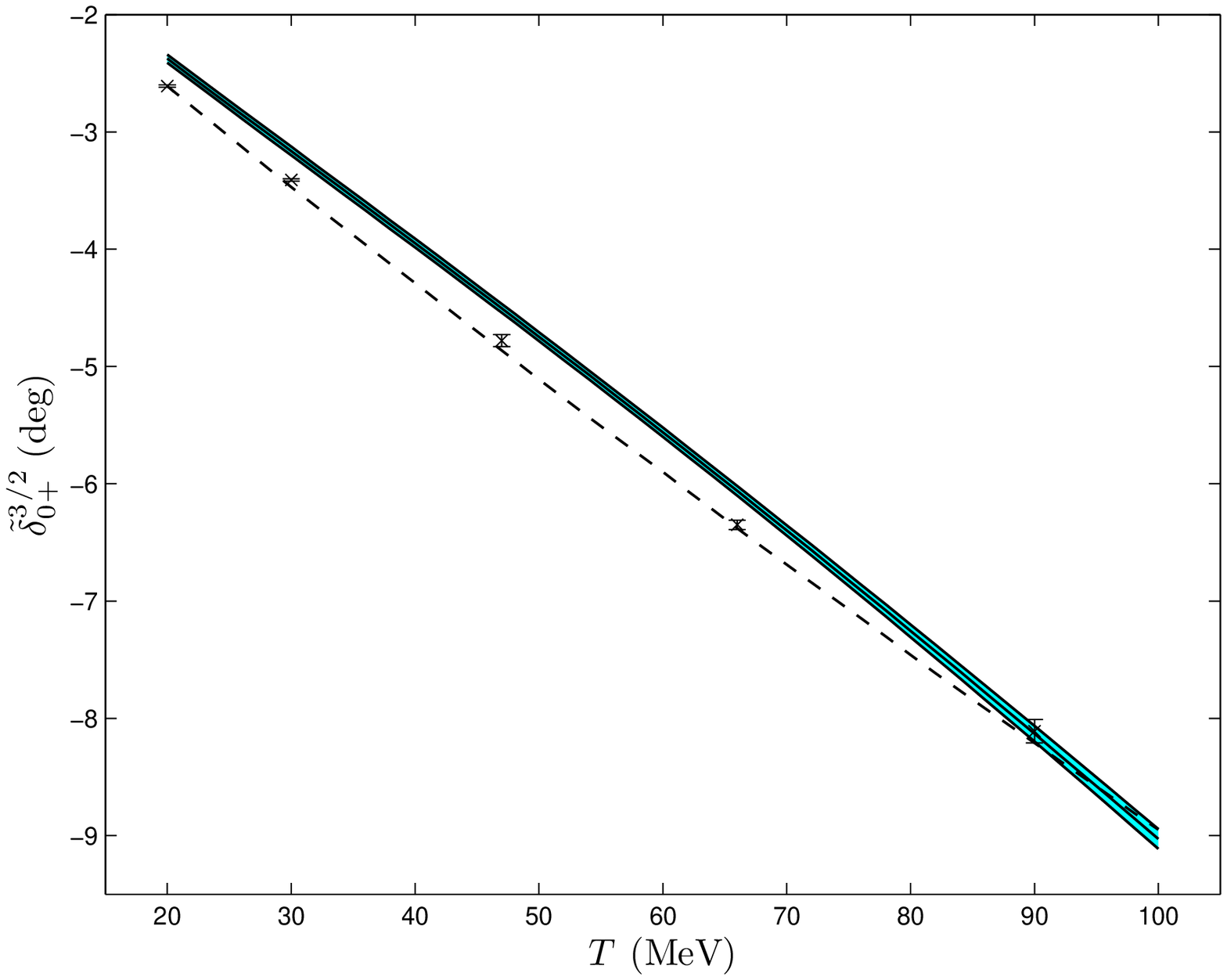}
\caption{\label{fig:a}The em-modified hadronic phase shift $\tilde{\delta}_{0+}^{3/2}$ (S31) from the present work (solid curve); the band around our 
solution indicates $1 \sigma$ uncertainties. Shown also is the current SAID solution (WI08) \cite{abws} (dashed curve), along with their five 
single-energy values (at $T=20$, $30$, $47$, $66$, and $90$ MeV).}
\end{center}
\end{figure}

\begin{figure}
\begin{center}
\includegraphics [width=15.5cm] {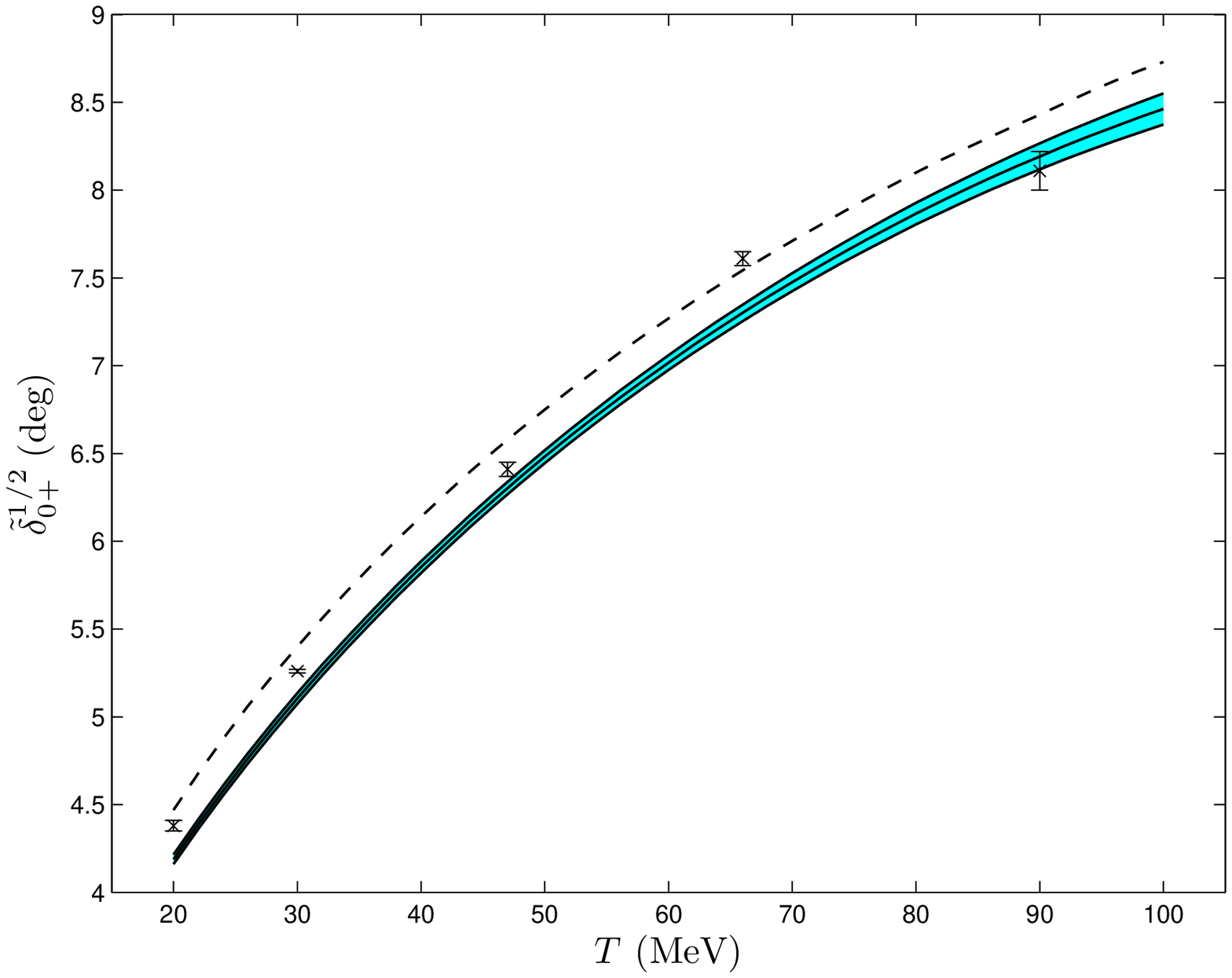}
\caption{\label{fig:b}The em-modified hadronic phase shift $\tilde{\delta}_{0+}^{1/2}$ (S11) from the present work (solid curve); the band around our 
solution indicates $1 \sigma$ uncertainties. Shown also is the current SAID solution (WI08) \cite{abws} (dashed curve), along with their five 
single-energy values (at $T=20$, $30$, $47$, $66$, and $90$ MeV).}
\end{center}
\end{figure}

\begin{figure}
\begin{center}
\includegraphics [width=15.5cm] {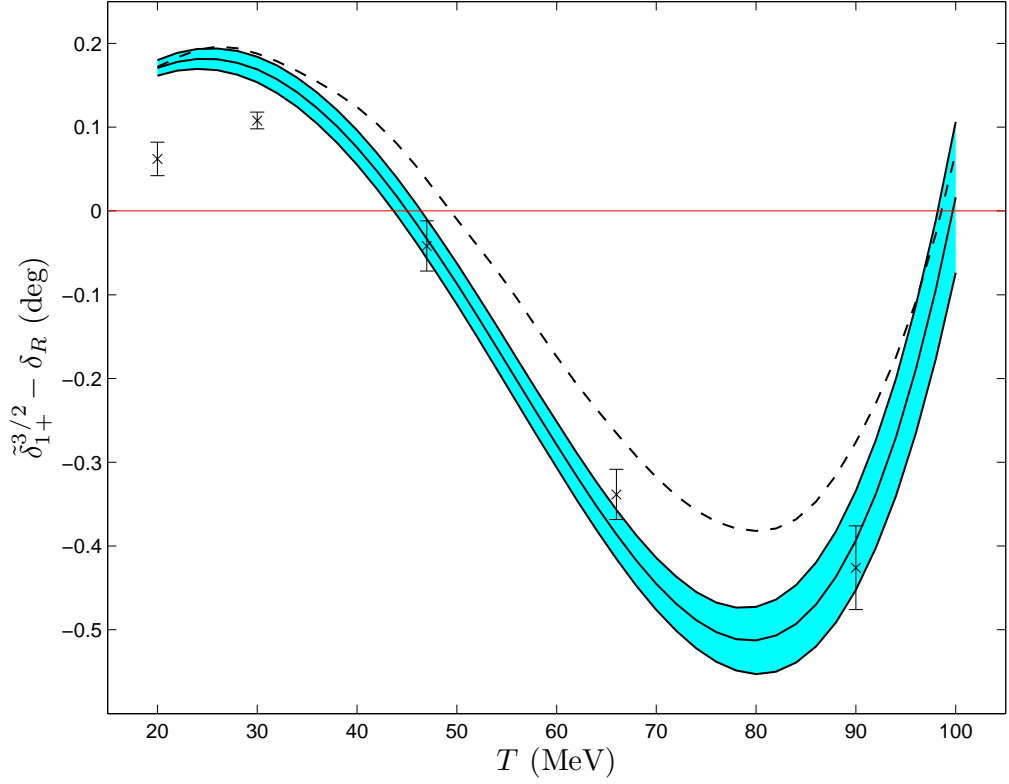}
\caption{\label{fig:c}The em-modified hadronic phase shift $\tilde{\delta}_{1+}^{3/2}$ (P33) from the present work (solid curve); the band around our solution indicates 
$1 \sigma$ uncertainties. Shown also is the current SAID solution (WI08) \cite{abws} (dashed curve), along with their five single-energy values (at $T=20$, $30$, $47$, 
$66$, and $90$ MeV). To enable the meaningful comparison of the values contained in this figure, an energy-dependent baseline $\delta_R$ ($=(0.20 \cdot T+1.54)T \cdot 10^{-2}$, 
with $T$ in MeV and $\delta_R(T)$ in degrees) was subtracted from all data.}
\end{center}
\end{figure}

\begin{figure}
\begin{center}
\includegraphics [width=15.5cm] {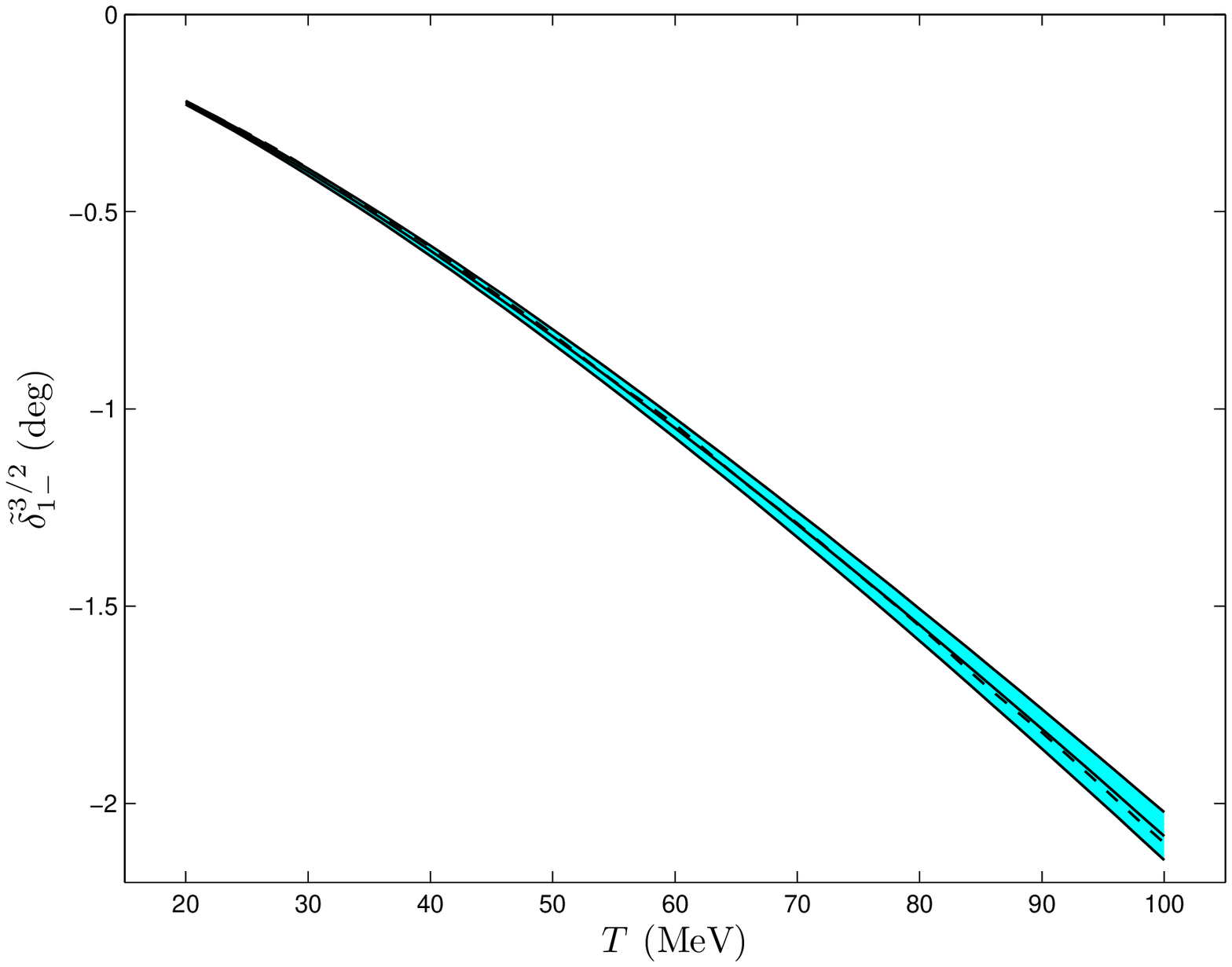}
\caption{\label{fig:d}The em-modified hadronic phase shift $\tilde{\delta}_{1-}^{3/2}$ (P31) from the present work (solid curve); the band around our 
solution indicates $1 \sigma$ uncertainties. Shown also is the current SAID solution (WI08) \cite{abws} (dashed curve).}
\end{center}
\end{figure}

\begin{figure}
\begin{center}
\includegraphics [width=15.5cm] {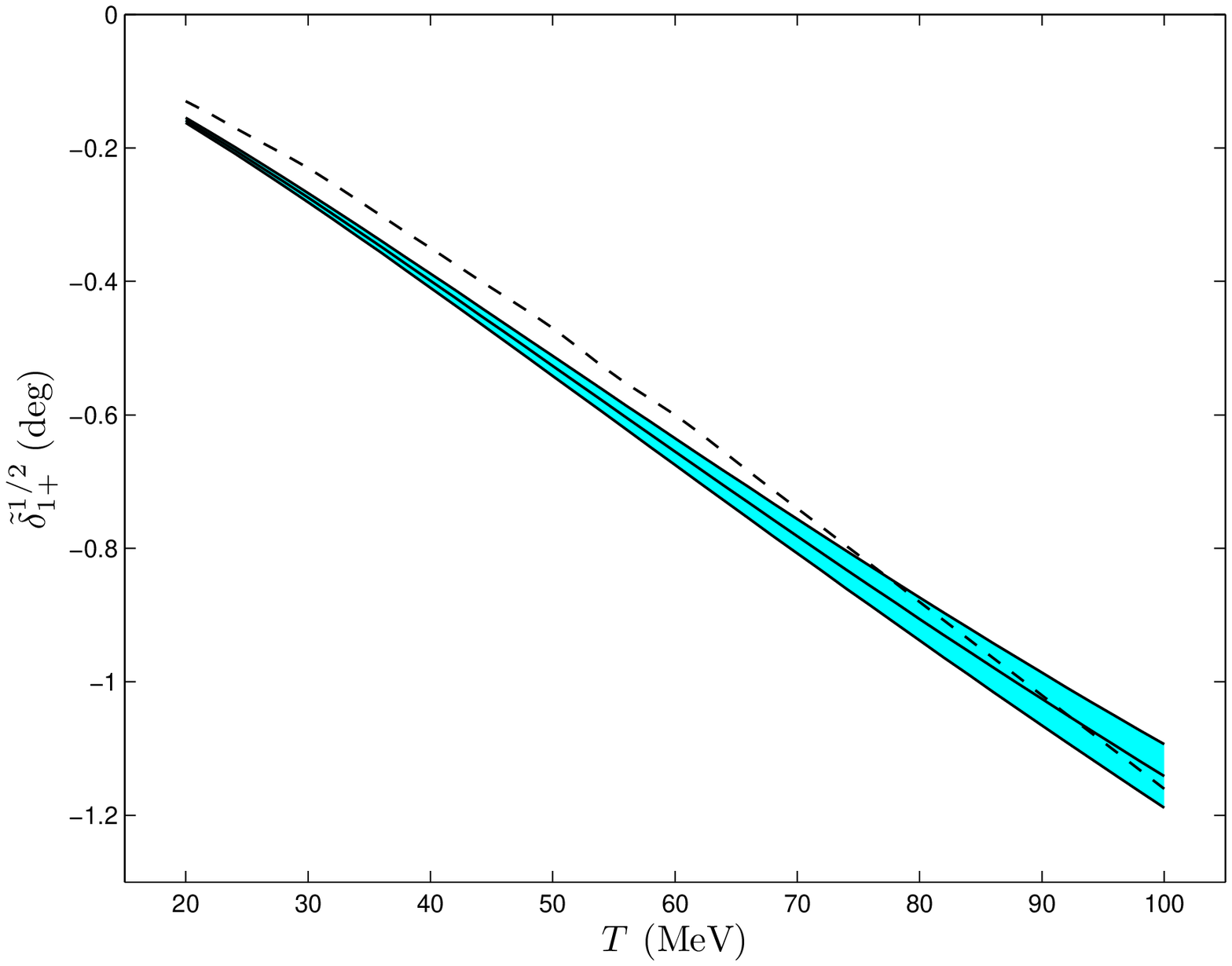}
\caption{\label{fig:e}The em-modified hadronic phase shift $\tilde{\delta}_{1+}^{1/2}$ (P13) from the present work (solid curve); the band around our 
solution indicates $1 \sigma$ uncertainties. Shown also is the current SAID solution (WI08) \cite{abws} (dashed curve).}
\end{center}
\end{figure}

\begin{figure}
\begin{center}
\includegraphics [width=15.5cm] {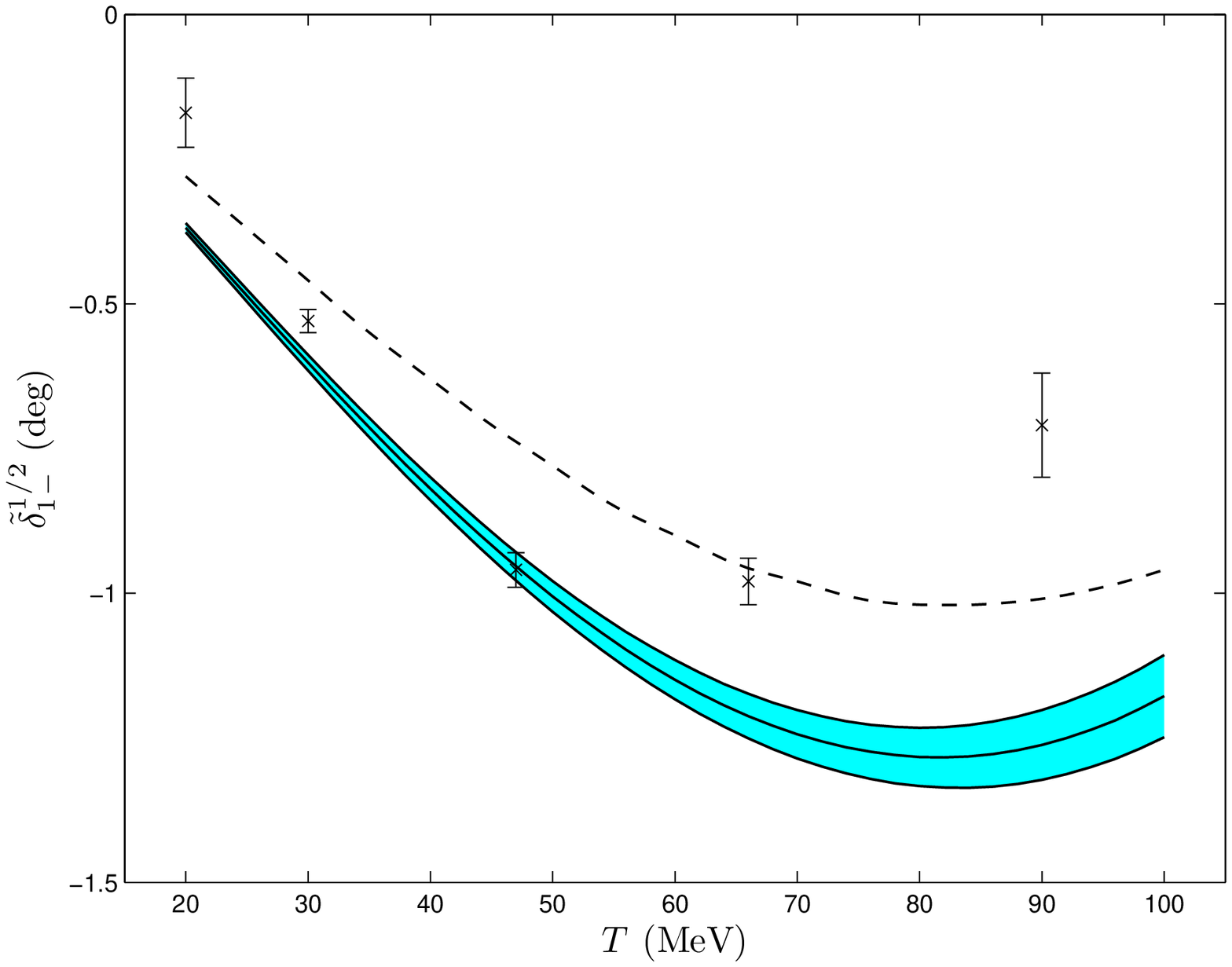}
\caption{\label{fig:f}The em-modified hadronic phase shift $\tilde{\delta}_{1-}^{1/2}$ (P11) from the present work (solid curve); the band around our 
solution indicates $1 \sigma$ uncertainties. Shown also is the current SAID solution (WI08) \cite{abws} (dashed curve), along with their five 
single-energy values (at $T=20$, $30$, $47$, $66$, and $90$ MeV).}
\end{center}
\end{figure}

The ETH model of Ref.~\cite{glmbg}, based on $s$- and $u$-channel diagrams with $N$ and $\Delta$ in the intermediate states, and $\sigma$ and $\rho$ $t$-channel exchanges, was subsequently fitted to the truncated combined $\pi^\pm p$ 
elastic-scattering databases. The model-parameter values showed reasonable stability when subjected to different criteria for the removal of data (see Table \ref{tab:ModelParameters}). Our result for the pseudovector $\pi N N$ 
coupling constant is $0.0726 \pm 0.0014$. On the basis of the results of the model fits, we obtained the em-modified hadronic scattering lengths and volumes (see Section \ref{sec:ModelConstants}), as well as the $s$- and $p$-wave 
em-modified hadronic phase shifts up to $T=100$ MeV (see Table \ref{tab:PhaseShifts} and Figs.~\ref{fig:a}-\ref{fig:f}). Large differences in the $s$-wave part of the interaction were found when comparing our hadronic phase shifts 
with the current SAID solution (WI08) \cite{abws} (see Figs.~\ref{fig:a} and \ref{fig:b}); there is general agreement in the $p$ waves, save for the em-modified hadronic phase shift $\tilde{\delta}_{1-}^{1/2}$.

\begin{figure}
\begin{center}
\includegraphics [width=15.5cm] {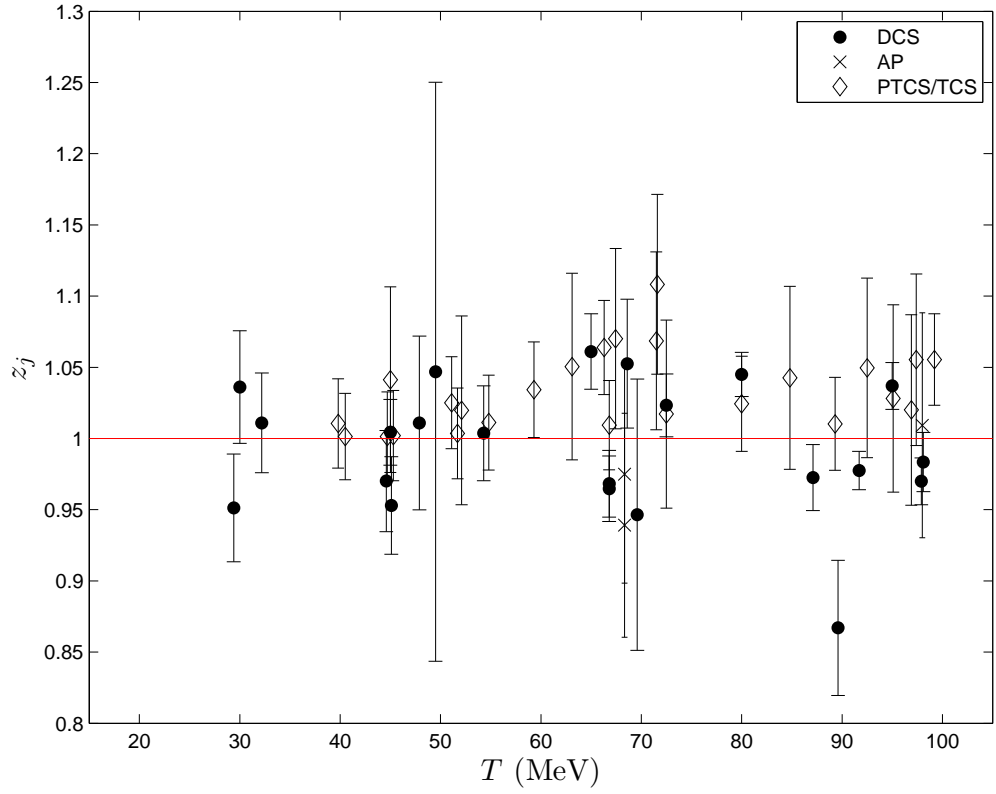}
\caption{\label{fig:sfpip}The scale factors $z_j$ for the $\pi^+ p$ data, obtained from the common fit to the truncated combined $\pi^\pm p$ elastic-scattering databases using the ETH model (see Section \ref{sec:Model}). 
The values, corresponding to the two data sets which were freely floated (see Table \ref{tab:DBpi+p}), have not been included; in the case of free floating, $(\chi_j^2)_{sc}=0$.}
\end{center}
\end{figure}

\begin{figure}
\begin{center}
\includegraphics [width=15.5cm] {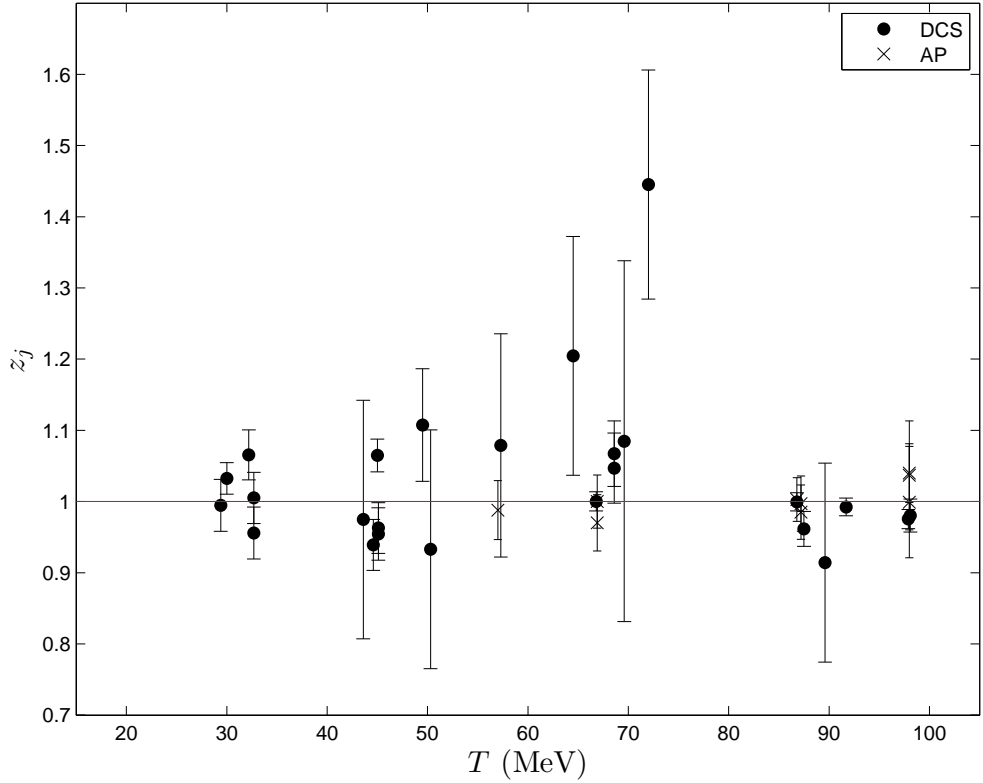}
\caption{\label{fig:sfpim}The scale factors $z_j$ for the $\pi^- p$ elastic-scattering data, obtained from the common fit to the truncated combined $\pi^\pm p$ elastic-scattering databases 
using the ETH model (see Section \ref{sec:Model}). The value, corresponding to the data set which was freely floated (see Table \ref{tab:DBpi-p}), has not been included; in the case of free 
floating, $(\chi_j^2)_{sc}=0$.}
\end{center}
\end{figure}

\begin{figure}
\begin{center}
\includegraphics [width=15.5cm] {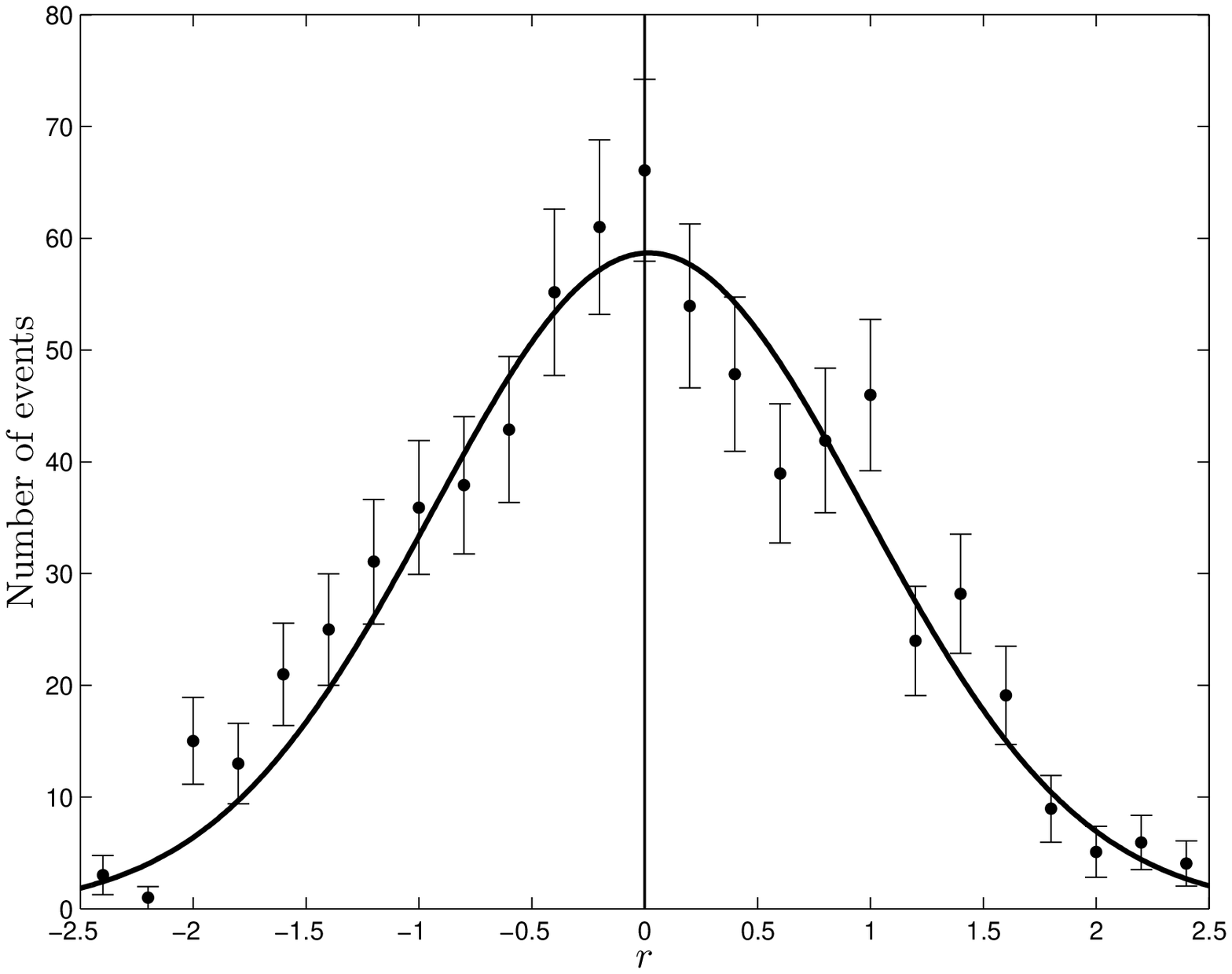}
\caption{\label{fig:residuals}The distribution of the normalised residuals, obtained from the common fit to the truncated combined $\pi^\pm p$ elastic-scattering databases using the ETH model 
(see Section \ref{sec:Model}). Also shown (solid curve) is the optimal Gaussian fit to the data.}
\end{center}
\end{figure}

Apart from analysing our results in terms of the assumed confidence level in the statistical tests, we also investigated the possibility of a bias in the analysis. To this end, we examined the energy dependence of 
the scale factors $z_j$, shown in Figs.~\ref{fig:sfpip} (for the $\pi^+ p$ data) and \ref{fig:sfpim} (for the $\pi^- p$ elastic-scattering data), as well as the characteristics of the distribution of the normalised residuals 
(Fig.~\ref{fig:residuals}). We did not find any significant deviation for these quantities from the expectations in a successful optimisation.

\begin{ack}
We are grateful to G.J. Wagner for drawing our attention to the statistical uncertainties of the scale factors $\hat{z}_j$ for free floating (see Eq.~(\ref{eq:zjoptunc})). We would like to thank W.S. Woolcock for his comments 
and suggestions.
\end{ack}

\newpage
\begin{table}[h!]
{\bf \caption{\label{tab:DBpi+p}}}The data sets comprising the truncated $\pi^+ p$ database, the pion laboratory kinetic energy $T$ (in MeV), the number of degrees of freedom $(NDF)_j$ for each data set, 
the scale factor $z_j$ which minimises $\chi_j^2$ of Eq.~(\ref{eq:chijsq}), the values of $(\chi_j^2)_{min}$, and the p-value of the fit for each data set. The numbers of this table correspond 
to the final fit to the data using the $K$-matrix parameterisations (see Section \ref{sec:K-Matrix_pi+p}).
\vspace{0.2cm}
\begin{center}
\begin{tabular}{|l|c|c|c|c|c|l|}
\hline
Data set & $T$ & $(NDF)_j$ & $z_j$ & $(\chi_j^2)_{min}$ & p-value & Comments \\
\hline
AULD79 & $47.90$ & $11$ & $1.0146$ & $15.6800$ & $0.1534$ & \\
RITCHIE83 & $65.00$ & $8$ & $1.0434$ & $17.2226$ & $0.0279$ & \\
RITCHIE83 & $72.50$ & $10$ & $1.0047$ & $4.7383$ & $0.9080$ & \\
RITCHIE83 & $80.00$ & $10$ & $1.0289$ & $19.0679$ & $0.0394$ & \\
RITCHIE83 & $95.00$ & $10$ & $1.0327$ & $13.1452$ & $0.2157$ & \\
FRANK83 & $29.40$ & $28$ & $1.0161$ & $19.4969$ & $0.8821$ & \\
FRANK83 & $49.50$ & $28$ & $1.0458$ & $33.4861$ & $0.2183$ & \\
FRANK83 & $69.60$ & $27$ & $0.9282$ & $23.6806$ & $0.6480$ & \\
FRANK83 & $89.60$ & $27$ & $0.8614$ & $29.2304$ & $0.3498$ & \\
BRACK86 & $66.80$ & $4$ & $0.8901$ & $2.4753$ & $0.6491$ & freely floated \\
BRACK86 & $86.80$ & $8$ & $0.9380$ & $15.9483$ & $0.0431$ & freely floated \\
BRACK86 & $91.70$ & $5$ & $0.9736$ & $11.6391$ & $0.0401$ & \\
BRACK86 & $97.90$ & $5$ & $0.9723$ & $7.2220$ & $0.2046$ & \\
BRACK88 & $66.80$ & $6$ & $0.9458$ & $11.5494$ & $0.0728$ & \\
BRACK88 & $66.80$ & $6$ & $0.9554$ & $10.1689$ & $0.1177$ & \\
WIEDNER89 & $54.30$ & $19$ & $0.9871$ & $14.7570$ & $0.7379$ & \\
BRACK90 & $30.00$ & $5$ & $1.0830$ & $10.6037$ & $0.0598$ & $79.40^\circ$ removed \\
BRACK90 & $45.00$ & $8$ & $1.0124$ & $7.6622$ & $0.4671$ & \\
BRACK95 & $87.10$ & $8$ & $0.9730$ & $13.8367$ & $0.0861$ & \\
BRACK95 & $98.10$ & $8$ & $0.9810$ & $14.8814$ & $0.0615$ & \\
JORAM95 & $45.10$ & $9$ & $0.9600$ & $20.1704$ & $0.0169$ & $124.42^\circ$ removed \\
JORAM95 & $68.60$ & $9$ & $1.0506$ & $8.2909$ & $0.5051$ & \\
JORAM95 & $32.20$ & $20$ & $1.0138$ & $33.5992$ & $0.0290$ & \\
JORAM95 & $44.60$ & $18$ & $0.9528$ & $29.9202$ & $0.0382$ & $30.74$, $35.40^\circ$ removed \\
SEVIOR89 & $98.00$ & $6$ & $1.0157$ & $5.4478$ & $0.4878$ & \\
\hline
\end{tabular}
\end{center}
\end{table}

\newpage
\begin{table*}
{\bf Table 1 continued}
\vspace{0.2cm}
\begin{center}
\begin{tabular}{|l|c|c|c|c|c|l|}
\hline
Data set & $T$ & $(NDF)_j$ & $z_j$ & $(\chi_j^2)_{min}$ & p-value & Comments \\
\hline
WIESER96 & $68.34$ & $3$ & $0.8899$ & $2.6924$ & $0.4415$ & \\
WIESER96 & $68.34$ & $4$ & $0.9202$ & $3.8288$ & $0.4297$ & \\
KRISS97 & $39.80$ & $1$ & $1.0129$ & $1.9961$ & $0.1577$ & \\
KRISS97 & $40.50$ & $1$ & $1.0019$ & $0.1775$ & $0.6735$ & \\
KRISS97 & $44.70$ & $1$ & $1.0027$ & $0.0768$ & $0.7817$ & \\
KRISS97 & $45.30$ & $1$ & $1.0034$ & $0.0946$ & $0.7584$ & \\
KRISS97 & $51.10$ & $1$ & $1.0246$ & $3.5130$ & $0.0609$ & \\
KRISS97 & $51.70$ & $1$ & $1.0029$ & $0.0588$ & $0.8084$ & \\
KRISS97 & $54.80$ & $1$ & $1.0077$ & $0.1738$ & $0.6768$ & \\
KRISS97 & $59.30$ & $1$ & $1.0256$ & $1.2984$ & $0.2545$ & \\
KRISS97 & $66.30$ & $2$ & $1.0497$ & $4.0101$ & $0.1347$ & \\
KRISS97 & $66.80$ & $2$ & $1.0074$ & $0.5779$ & $0.7491$ & \\
KRISS97 & $80.00$ & $1$ & $1.0136$ & $0.3366$ & $0.5618$ & \\
KRISS97 & $89.30$ & $1$ & $1.0078$ & $0.2795$ & $0.5970$ & \\
KRISS97 & $99.20$ & $1$ & $1.0560$ & $4.2590$ & $0.0390$ & \\
FRIEDMAN99 & $45.00$ & $1$ & $1.0437$ & $2.3178$ & $0.1279$ & \\
FRIEDMAN99 & $52.10$ & $1$ & $1.0182$ & $0.2772$ & $0.5986$ & \\
FRIEDMAN99 & $63.10$ & $1$ & $1.0363$ & $0.4904$ & $0.4838$ & \\
FRIEDMAN99 & $67.45$ & $2$ & $1.0517$ & $1.2397$ & $0.5380$ & \\
FRIEDMAN99 & $71.50$ & $2$ & $1.0490$ & $0.8114$ & $0.6665$ & \\
FRIEDMAN99 & $92.50$ & $2$ & $1.0429$ & $0.5872$ & $0.7456$ & \\
CARTER71 & $71.60$ & $1$ & $1.0921$ & $2.6734$ & $0.1020$ & \\
CARTER71 & $97.40$ & $1$ & $1.0498$ & $0.6952$ & $0.4044$ & \\
PEDRONI78 & $72.50$ & $1$ & $1.0121$ & $0.1329$ & $0.7155$ & \\
PEDRONI78 & $84.80$ & $1$ & $1.0311$ & $0.3258$ & $0.5682$ & \\
PEDRONI78 & $95.10$ & $1$ & $1.0230$ & $0.2030$ & $0.6523$ & \\
PEDRONI78 & $96.90$ & $1$ & $1.0167$ & $0.1330$ & $0.7153$ & \\
\hline
\end{tabular}
\end{center}
\end{table*}

\newpage
\begin{table}
{\bf \caption{\label{tab:DBpi-p}}}The data sets comprising the truncated $\pi^- p$ elastic-scattering database, the pion laboratory kinetic energy $T$ (in MeV), the number of degrees of freedom $(NDF)_j$ for each data set, 
the scale factor $z_j$ which minimises $\chi_j^2$ of Eq.~(\ref{eq:chijsq}), the values of $(\chi_j^2)_{min}$, and the p-value of the fit for each data set. The numbers of this table correspond 
to the final fit to the data using the $K$-matrix parameterisations (see Section \ref{sec:K-Matrix_pi-p}).
\vspace{0.2cm}
\begin{center}
\begin{tabular}{|l|c|c|c|c|c|l|}
\hline
Data set & $T$ & $(NDF)_j$ & $z_j$ & $(\chi_j^2)_{min}$ & p-value & Comments \\
\hline
FRANK83 & $29.40$ & $28$ & $0.9832$ & $30.9926$ & $0.3174$ & \\
FRANK83 & $49.50$ & $28$ & $1.1007$ & $30.1075$ & $0.3581$ & \\
FRANK83 & $69.60$ & $27$ & $1.0953$ & $25.6707$ & $0.5369$ & \\
FRANK83 & $89.60$ & $27$ & $0.9479$ & $25.4255$ & $0.5506$ & \\
BRACK86 & $66.80$ & $5$ & $0.9973$ & $13.9690$ & $0.0158$ & \\
BRACK86 & $86.80$ & $5$ & $1.0036$ & $1.4172$ & $0.9224$ & \\
BRACK86 & $91.70$ & $5$ & $0.9963$ & $2.8898$ & $0.7170$ & \\
BRACK86 & $97.90$ & $5$ & $1.0002$ & $5.9408$ & $0.3120$ & \\
WIEDNER89 & $54.30$ & $18$ & $1.1597$ & $23.6926$ & $0.1654$ & $15.55^\circ$ removed, freely floated \\
BRACK90 & $30.00$ & $5$ & $1.0204$ & $4.8808$ & $0.4306$ & \\
BRACK90 & $45.00$ & $9$ & $1.0536$ & $12.3031$ & $0.1968$ & \\
BRACK95 & $87.50$ & $6$ & $0.9830$ & $10.3543$ & $0.1105$ & \\
BRACK95 & $98.10$ & $7$ & $1.0092$ & $8.2223$ & $0.3134$ & $36.70^\circ$ removed \\
JORAM95 & $32.70$ & $4$ & $0.9951$ & $4.0883$ & $0.3942$ & \\
JORAM95 & $32.70$ & $2$ & $0.9527$ & $5.7974$ & $0.0551$ & \\
JORAM95 & $45.10$ & $4$ & $0.9561$ & $12.4590$ & $0.0142$ & \\
JORAM95 & $45.10$ & $3$ & $0.9462$ & $9.2581$ & $0.0260$ & \\
JORAM95 & $68.60$ & $7$ & $1.0863$ & $14.2673$ & $0.0466$ & \\
JORAM95 & $68.60$ & $3$ & $1.0314$ & $2.2747$ & $0.5174$ & \\
JORAM95 & $32.20$ & $20$ & $1.0617$ & $21.3392$ & $0.3774$ & \\
JORAM95 & $44.60$ & $20$ & $0.9462$ & $29.7408$ & $0.0742$ & \\
JANOUSCH97 & $43.60$ & $1$ & $1.0420$ & $0.1682$ & $0.6817$ & \\
JANOUSCH97 & $50.30$ & $1$ & $1.0364$ & $0.1557$ & $0.6931$ & \\
JANOUSCH97 & $57.30$ & $1$ & $1.0830$ & $4.5370$ & $0.0332$ & \\
\hline
\end{tabular}
\end{center}
\end{table}

\newpage
\begin{table*}
{\bf Table 2 continued}
\vspace{0.2cm}
\begin{center}
\begin{tabular}{|l|c|c|c|c|c|l|}
\hline
Data set & $T$ & $(NDF)_j$ & $z_j$ & $(\chi_j^2)_{min}$ & p-value & Comments \\
\hline
JANOUSCH97 & $64.50$ & $1$ & $0.9962$ & $0.0010$ & $0.9753$ & \\
JANOUSCH97 & $72.00$ & $1$ & $1.3045$ & $4.8348$ & $0.0279$ & \\
ALDER83 & $98.00$ & $6$ & $1.0335$ & $5.0919$ & $0.5321$ & \\
SEVIOR89 & $98.00$ & $5$ & $0.9882$ & $1.5869$ & $0.9028$ & \\
HOFMAN98 & $86.80$ & $11$ & $1.0020$ & $5.8362$ & $0.8841$ & \\
PATTERSON02 & $57.00$ & $10$ & $0.9365$ & $11.0276$ & $0.3554$ & \\
PATTERSON02 & $66.90$ & $9$ & $0.9985$ & $4.3343$ & $0.8881$ & \\
PATTERSON02 & $66.90$ & $10$ & $0.9479$ & $18.0250$ & $0.0545$ & \\
PATTERSON02 & $87.20$ & $11$ & $0.9835$ & $8.2469$ & $0.6910$ & \\
PATTERSON02 & $87.20$ & $11$ & $0.9945$ & $5.0493$ & $0.9288$ & \\
PATTERSON02 & $98.00$ & $12$ & $0.9954$ & $6.9731$ & $0.8594$ & \\
\hline
\end{tabular}
\end{center}
\end{table*}

\vspace{0.5cm}
\begin{table}
{\bf \caption{\label{tab:ModelParameters}}}The values of the seven parameters of the ETH model obtained from the common fits to the truncated combined $\pi^\pm p$ elastic-scattering databases 
for three values of $\mathrm{p}_{min}$ (the confidence level in the statistical tests); these three $\mathrm{p}_{min}$ values correspond to a $3$, $2.5$, and $2 \sigma$ effect in the normal 
distribution, respectively. The uncertainties correspond to the fit for $\mathrm{p}_{min} \approx 1.24 \cdot 10^{-2}$.
\vspace{0.2cm}
\begin{center}
\begin{tabular}{|l|c|c|c|c|c|}
\hline
 & $\mathrm{p}_{min} \approx 2.70 \cdot 10^{-3}$ & $\mathrm{p}_{min} \approx 1.24 \cdot 10^{-2}$ & $\mathrm{p}_{min} \approx 4.55 \cdot 10^{-2}$ & uncertainty \\
\hline
$G_{\sigma}(GeV^{-2})$ & $27.43$ & $27.48$ & $27.37$ & $0.86$ \\
$K_{\sigma}$ & $0.014$ & $0.016$ & $0.075$ & $0.034$ \\
$G_{\rho}(GeV^{-2})$ & $54.71$ & $54.67$ & $55.98$ & $0.61$ \\
$K_{\rho}$ & $0.66$ & $0.66$ & $1.35$ & $0.41$ \\
$g_{\pi NN}$ & $12.84$ & $12.84$ & $13.07$ & $0.12$ \\
$g_{\pi N \Delta}$ & $29.78$ & $29.77$ & $29.31$ & $0.26$ \\
$Z$ & $-0.550$ & $-0.552$ & $-0.439$ & $0.056$ \\
\hline
\end{tabular}
\end{center}
\end{table}

\newpage
\begin{table}
{\bf \caption{\label{tab:HessianMatrix}}}The correlation matrix for the seven parameters of the ETH model for the common fit to the truncated combined $\pi^\pm p$ elastic-scattering databases 
for $\mathrm{p}_{min} \approx 1.24 \cdot 10^{-2}$.
\vspace{0.2cm}
\begin{center}
\begin{tabular}{|l|c|c|c|c|c|c|c|}
\hline
 & $G_{\sigma}$ & $K_{\sigma}$ & $G_{\rho}$ & $K_{\rho}$ & $g_{\pi NN}$ & $g_{\pi N \Delta}$ &$Z$ \\
\hline
$G_{\sigma}$ & $1.0000$ & $0.5095$ & $-0.0886$ & $-0.0378$ & $0.1030$ & $-0.1602$ & $-0.1999$ \\
$K_{\sigma}$ & $0.5095$ & $1.0000$ & $0.7314$ & $0.7974$ & $0.8847$ & $-0.9210$ & $0.7176$ \\
$G_{\rho}$ & $-0.0886$ & $0.7314$ & $1.0000$ & $0.9044$ & $0.9038$ & $-0.8487$ & $0.8977$ \\
$K_{\rho}$ & $-0.0378$ & $0.7974$ & $0.9044$ & $1.0000$ & $0.9522$ & $-0.9284$ & $0.9530$ \\
$g_{\pi NN}$ & $0.1030$ & $0.8847$ & $0.9038$ & $0.9522$ & $1.0000$ & $-0.9497$ & $0.9216$ \\
$g_{\pi N \Delta}$ & $-0.1602$ & $-0.9210$ & $-0.8487$ & $-0.9284$ & $-0.9497$ & $1.0000$ & $-0.9001$ \\
$Z$ & $-0.1999$ & $0.7176$ & $0.8977$ & $0.9530$ & $0.9216$ & $-0.9001$ & $1.0000$ \\
\hline
\end{tabular}
\end{center}
\end{table}
\vspace{0.5cm}

\newpage
\begin{table}
{\bf \caption{\label{tab:PhaseShifts}}}The values of the six $s$- and $p$-wave em-modified hadronic phase shifts (in degrees), obtained on the basis of the results of Tables \ref{tab:ModelParameters} (for $\mathrm{p}_{min} \approx 1.24 \cdot 10^{-2}$) and \ref{tab:HessianMatrix}.
\vspace{0.2cm}
\begin{center}
\begin{tabular}{|c|c|c|c|c|c|c|}
\hline
$T(MeV)$ & $\tilde{\delta}_{0+}^{3/2}$ (S31) & $\tilde{\delta}_{0+}^{1/2}$ (S11) & $\tilde{\delta}_{1+}^{3/2}$ (P33) & $\tilde{\delta}_{1-}^{3/2}$ (P31) & $\tilde{\delta}_{1+}^{1/2}$ (P13) & $\tilde{\delta}_{1-}^{1/2}$ (P11) \\ 
\hline
$20$ & $-2.375(34)$ & $4.189(27)$ & $1.2787(94)$ & $-0.2239(45)$ & $-0.1588(37)$ & $-0.3687(80)$ \\
$25$ & $-2.772(36)$ & $4.673(29)$ & $1.817(12)$ & $-0.3083(63)$ & $-0.2153(52)$ & $-0.486(11)$ \\
$30$ & $-3.164(37)$ & $5.105(30)$ & $2.431(15)$ & $-0.3996(83)$ & $-0.2747(68)$ & $-0.602(14)$ \\
$35$ & $-3.555(37)$ & $5.496(31)$ & $3.122(18)$ & $-0.497(11)$ & $-0.3361(86)$ & $-0.714(17)$ \\
$40$ & $-3.949(37)$ & $5.852(33)$ & $3.892(21)$ & $-0.599(13)$ & $-0.399(11)$ & $-0.820(20)$ \\
$45$ & $-4.345(37)$ & $6.180(34)$ & $4.744(23)$ & $-0.706(16)$ & $-0.463(13)$ & $-0.918(23)$ \\
$50$ & $-4.746(37)$ & $6.482(37)$ & $5.683(24)$ & $-0.816(18)$ & $-0.527(15)$ & $-1.006(27)$ \\
$55$ & $-5.151(38)$ & $6.760(39)$ & $6.715(26)$ & $-0.931(21)$ & $-0.591(17)$ & $-1.084(30)$ \\
$60$ & $-5.561(39)$ & $7.018(42)$ & $7.845(27)$ & $-1.049(25)$ & $-0.655(20)$ & $-1.150(34)$ \\
$65$ & $-5.977(41)$ & $7.256(46)$ & $9.081(29)$ & $-1.169(28)$ & $-0.719(23)$ & $-1.204(38)$ \\
$70$ & $-6.397(44)$ & $7.476(51)$ & $10.433(31)$ & $-1.293(32)$ & $-0.782(26)$ & $-1.244(42)$ \\
$75$ & $-6.823(48)$ & $7.679(56)$ & $11.909(35)$ & $-1.419(36)$ & $-0.844(29)$ & $-1.271(46)$ \\
$80$ & $-7.254(53)$ & $7.865(61)$ & $13.519(40)$ & $-1.547(41)$ & $-0.906(32)$ & $-1.283(50)$ \\
$85$ & $-7.690(59)$ & $8.036(67)$ & $15.277(48)$ & $-1.678(45)$ & $-0.966(36)$ & $-1.281(55)$ \\
$90$ & $-8.131(67)$ & $8.192(74)$ & $17.193(59)$ & $-1.811(50)$ & $-1.026(39)$ & $-1.263(60)$ \\
$95$ & $-8.577(75)$ & $8.334(81)$ & $19.282(73)$ & $-1.946(55)$ & $-1.084(43)$ & $-1.229(66)$ \\
$100$ & $-9.028(85)$ & $8.462(89)$ & $21.556(90)$ & $-2.083(61)$ & $-1.141(48)$ & $-1.178(71)$ \\
\hline
\end{tabular}
\end{center}
\end{table}

\clearpage

\clearpage

\clearpage

\clearpage

\clearpage

\clearpage

\clearpage

\clearpage

\begin{thebibliography}{99}
\bibitem{mworg} E. Matsinos, W.S. Woolcock, G.C. Oades, G. Rasche, A. Gashi, Nucl. Phys. A 778 (2006) 95-123.
\bibitem{mrw2} E. Matsinos, G. Rasche, `Analysis of the low-energy $\pi^\pm p$ elastic-scattering data of the CHAOS Collaboration', submitted to Nucl. Phys. A; http://arxiv.org/abs/1203.3635.
\bibitem{denz} H. Denz \etal, Phys. Lett. B 633 (2006) 209-13; the values of the $\pi^\pm p$ elastic-scattering cross section are available from the site: http://tobias-lib.uni-tuebingen.de/dbt/volltexte/2004/1323/.
\bibitem{mrw3} E. Matsinos, G. Rasche, `Analysis of the low-energy $\pi^- p$ charge-exchange data', in preparation; a preliminary version of this paper may be found in http://arxiv.org/abs/1203.3856.
\bibitem{gmorw} A. Gashi, E. Matsinos, G.C. Oades, G. Rasche, W.S. Woolcock, Nucl. Phys. A 686 (2001) 447-62; A. Gashi, E. Matsinos, G.C. Oades, G. Rasche, W.S. Woolcock, {\it ibid.} pp.~463-77.
\bibitem{pdg} K. Nakamura \etal~(Particle Data Group), J. Phys. G 37 (2010) 075021.
\bibitem{glk} W.R. Gibbs, Li Ai, W.B. Kaufmann, Phys. Rev. Lett. 74 (1995) 3740-3.
\bibitem{m} E. Matsinos, Phys. Rev. C 56 (1997) 3014-25.
\bibitem{ar} R.A. Arndt, L.D. Roper, Nucl. Phys. B 50 (1972) 285-300.
\bibitem{ega} E.G. Auld \etal, Can. J. Phys. 57 (1979) 73-8.
\bibitem{bgr} B.G. Ritchie \etal, Phys. Lett. B 125 (1983) 128-32.
\bibitem{jsf} J.S. Frank \etal, Phys. Rev. D 28 (1983) 1569-85.
\bibitem{jtb} J.T. Brack \etal, Phys. Rev. C 34 (1986) 1771-8.
\bibitem{jtbb} J.T. Brack \etal, Phys. Rev. C 38 (1988) 2427-9.
\bibitem{uw} U. Wiedner \etal, Phys. Rev. D 40 (1989) 3568-81.
\bibitem{jtbbb} J.T. Brack \etal, Phys. Rev. C 41 (1990) 2202-14.
\bibitem{jtbbbb} J.T. Brack \etal, Phys. Rev. C 51 (1995) 929-36.
\bibitem{cj} Ch. Joram \etal, Phys. Rev. C 51 (1995) 2144-58; Ch. Joram \etal, {\it ibid.} pp.~2159-65.
\bibitem{mes} M.E. Sevior \etal, Phys. Rev. C 40 (1989) 2780-8.
\bibitem{rw} R. Wieser \etal, Phys. Rev. C 54 (1996) 1930-4.
\bibitem{bjk} B.J. Kriss \etal, $\pi N$ Newsletter 12 (1997) 20-5; B.J. Kriss \etal, Phys. Rev. C 59 (1999) 1480-7.
\bibitem{ef} E. Friedman, $\pi N$ Newsletter 15 (1999) 37-42.
\bibitem{cwbbd} A.A. Carter, J.R. Williams, D.V. Bugg, P.J. Bussey, D.R. Dance, Nucl. Phys. B 26 (1971) 445-60.
\bibitem{ep} E. Pedroni \etal, Nucl. Phys. A 300 (1978) 321-47.
\bibitem{mj} M. Janousch \etal, Phys. Lett. B 414 (1997) 237-41.
\bibitem{jca} J.C. Alder \etal, Phys. Rev. D 27 (1983) 1040-55.
\bibitem{gjh} G.J. Hofman \etal, Phys. Rev. C 58 (1998) 3484-93.
\bibitem{jdp} J.D. Patterson \etal, Phys. Rev. C 66 (2002) 025207.
\bibitem{brt} P.Y. Bertin \etal, Nucl. Phys. B 106 (1976) 341-54.
\bibitem{bussey} P.J. Bussey \etal, Nucl. Phys. B 58 (1973) 363-77.
\bibitem{meier} R. Meier \etal, Phys. Lett. B 588 (2004) 155-62.
\bibitem{ss} H.-Ch. Schr{\"o}der {\it et al.}, Eur. Phys. J. C 21 (2001) 473-88.
\bibitem{orwmg} G.C. Oades, G. Rasche, W.S. Woolcock, E. Matsinos, A. Gashi, Nucl. Phys. A 794 (2007) 73-86.
\bibitem{abws} R.A. Arndt, W.J. Briscoe, I.I. Strakovsky, R.L. Workman, Phys. Rev. C 74 (2006) 045205; SAID PSA Tool, available at http://gwdac.phys.gwu.edu.
\bibitem{jms} F. James, `MINUIT - Function Minimization and Error Analysis', CERN Program Library Long Writeup D506.
\bibitem{fm} N. Fettes, E. Matsinos, Phys. Rev. C 55 (1997) 464-73.
\bibitem{ew} T.E.O. Ericson, W. Weise, Pions and Nuclei, Clarendon Press, Oxford, 1988.
\bibitem{glmbg} P.F.A. Goudsmit, H.J. Leisi, E. Matsinos, B.L. Birbrair, A.B. Gridnev, Nucl. Phys. A 575 (1994) 673-706; additional references on the development of the ETH model may be found at http://people.web.psi.ch/matsinos/0\_home.htm.
\bibitem{kp} R. Koch, E. Pietarinen, Nucl. Phys. A 336 (1980) 331-46; R. Koch, Nucl. Phys. A 448 (1986) 707-31.
\bibitem{aco} J.M. Alacr{\'o}n, J.M. Camalich, J.A. Oller, Phys. Rev. D 85 (2012) 051503.
\end{thebibliography}
\end{document}